\documentclass{emulateapj}
\usepackage{graphicx,natbib}
\usepackage{bm}
\usepackage{epsfig}
\shorttitle{Short gamma-ray bursts as standard sirens}
\shortauthors{Nissanke et al.}

\begin{document}

\title{Exploring short gamma-ray bursts as gravitational-wave standard sirens}

\author{Samaya Nissanke\altaffilmark{1,2}, Daniel E.\ Holz\altaffilmark{3},  Scott A.\
  Hughes\altaffilmark{2}, Neal
  Dalal\altaffilmark{1}, Jonathan L.\ Sievers\altaffilmark{1}}

\altaffiltext{1}{CITA, University of Toronto, 60 St.~George St.,
  Toronto, ON, M5S 3H8, Canada}

\altaffiltext{2}{Department of Physics and MIT Kavli Institute, 77
  Massachusetts Avenue, Cambridge, MA 02139}

\altaffiltext{3}{Theoretical Division, Los Alamos National Laboratory,
  Los Alamos, NM 87545}

\begin{abstract}

Recent observations support the hypothesis that a large fraction of
``short-hard'' gamma-ray bursts (SHBs) are associated with the
inspiral and merger of compact binaries.  Since gravitational-wave
(GW) measurements of well-localized inspiraling binaries can measure
absolute source distances, simultaneous observation of a binary's GWs
and SHB would allow us to directly and independently determine both
the binary's luminosity distance and its redshift.  Such a ``standard
siren'' (the GW analog of a standard candle) would provide an
excellent probe of the nearby ($z \lesssim 0.3$) universe's
expansion, independent of the cosmological distance ladder,
thereby complementing other standard candles.  Previous work explored this
idea using a simplified formalism to study measurement by advanced GW
detector networks, incorporating a high signal-to-noise ratio limit to
describe the probability distribution for measured parameters.  In
this paper we eliminate this simplification, constructing
distributions with a Markov-Chain Monte-Carlo technique.  We assume
that each SHB observation gives source sky position and time of
coalescence, and we take non-spinning binary neutron star and black
hole-neutron star coalescences as plausible SHB progenitors.  We
examine how well parameters (particularly distance) can be measured
from GW observations of SHBs by a range of ground-based detector
networks.  We find that earlier estimates overstate how well distances
can be measured, even at fairly large signal-to-noise ratio.  The
fundamental limitation to determining distance proves to be a
degeneracy between distance and source inclination.  Overcoming this
limitation requires that we either break this degeneracy, or measure
enough sources to broadly sample the inclination distribution.
\end{abstract}


\section{Introduction}

\subsection{Overview}

Two multi-kilometer interferometric gravitational-wave (GW) detectors
are presently in operation: LIGO\footnote{http://www.ligo.caltech.edu}
and Virgo\footnote{http://www.virgo.infn.it}. They are sensitive to
GWs produced by the coalescence of two neutron stars to a distance of
roughly 30 Mpc, and to the coalescence of a neutron star with a
$10M_{\odot}$ black hole to roughly 60 Mpc.  Over the next several
years, these detectors will undergo upgrades which are expected to
extend their range by a factor $\sim 10$.  Most estimates suggest that
detectors at advanced sensitivity should measure at least a few, and
possibly a few dozen, binary coalescences every year (e.g.,
\citealt{koppa08,ligo_rates}).

It has long been argued that neutron star-neutron star (NS-NS) and
neutron star-black hole (NS-BH) mergers are likely to be accompanied
by a gamma-ray burst \citep{eichler89}.  Recent evidence supports the
hypothesis that many short-hard gamma-ray bursts (SHBs) are indeed
associated with such mergers (\citealt{fox05}, \citealt{nakar06},
\citealt{berger07}, \citealt{perleyetal08}).  This suggests that it
may be possible to simultaneously measure a binary coalescence in
gamma rays (and associated afterglow emission) and in GWs
\citep{dietz09}.  The combined electromagnetic and gravitational view
of these objects will teach us substantially more than what we learn
from either data channel alone.  Because GWs track a system's global
mass and energy dynamics, it has long been known that measuring GWs from a coalescing binary allows us to determine, in the
ideal case, ``intrinsic'' binary properties such as the masses and spins
of its members with exquisite
accuracy (\citealt{fc93}, \citealt{cf94}).  As we describe in
the following subsection, it has also long been appreciated that GWs
can determine a system's ``extrinsic'' properties
\citep{schutz86} such as location on the sky and distance to the
source.  In particular, the amplitude of a binary's GWs directly
encodes its luminosity distance.  Direct measurement of a coalescing
binary could thus be used as a cosmic distance measure: Binary
inspiral would be a ``standard siren'' (the GW equivalent of a
standard candle, so-called due to the sound-like nature of GWs) whose
calibration depends only on the validity of general
relativity~\citep{HolzHughes05,dalaletal}.

Unfortunately, GWs alone do not measure extrinsic parameters as
accurately as the intrinsic ones. As we describe in more detail in the
following section, GW observation of a binary measures a complicated
combination of its distance, its position on the sky, and its
orientation, with overall fractional accuracy $\sim
1/\mbox{signal-to-noise}$.  As distance is degenerate with these
angles, using GWs to measure absolute distance to a source requires a
mechanism to break the degeneracy. Associating the GW coalescence
waves with a short-hard gamma-ray burst (SHB) is a near-perfect way to
break some of these degeneracies.

In this paper we explore the ability of the near-future advanced
LIGO-Virgo detector network to constrain binary parameters (especially
distance), when used in conjunction with electromagnetic observations
of the same event (such as an associated SHB). We also examine how
well these measurements can be improved if planned detectors in
Western Australia (AIGO\footnote{http://www.gravity.uwa.edu.au}) and
in Japan's Kamioka mine
(LCGT\footnote{http://gw.icrr.u-tokyo.ac.jp:8888/lcgt/}) are
operational. This paper substantially updates and improves upon
earlier work \citep[hereafter DHHJ06]{dalaletal}, using a more
sophisticated parameter estimation technique. In the next section we
review standard sirens, and in Sec.\ {\ref{sec:dalaletal}} we briefly
summarize DHHJ06. The next subsection describes the
organization and background relevant for the rest of the paper.

\subsection{Standard sirens}
\label{sec:sirens}

It has long been recognized that GW inspiral measurements could be
used as powerful tools for cosmology.  {\cite{schutz86}} first
demonstrated this by analyzing how binary coalescences allow a direct
measurement of the Hubble constant; {\cite{markovic93}} and
{\cite{fc93}} subsequently generalized this approach to include other
cosmological parameters. More recently, there has been much interest
in the measurements enabled when GWs from a merger are accompanied by
a counterpart in the electromagnetic spectrum~(\citealt{bloometal09},
\citealt{phinney09}, \citealt{kk09}).  In this paper we focus
exclusively on GW observations of binaries that have an independent
sky position furnished by electromagnetic observations.

We begin by examining gravitational waves from binary inspiral as
measured in a single detector.  We only present here the lowest order
contribution to the waves; in subsequent calculations our results are
taken to higher order (see Sec.\ \ref{sec:gwform}).  The leading
waveform generated by a source at luminosity distance $D_L$,
corresponding to redshift $z$, is given by
\begin{eqnarray}
h_+ &=& \frac{2(1 + z){\cal M}}{D_L}\left[\pi(1 + z){\cal
M}f\right]^{2/3}\times\nonumber\\
& & \qquad\qquad\qquad\left(1 + \cos^2\iota\right)\cos2\Phi_N(t)\;,
\nonumber\\
h_\times &=& -\frac{4(1 + z){\cal M}}{D_L}\left[\pi(1 + z){\cal
M}f\right]^{2/3}\cos\iota \sin2\Phi_N(t)\;,
\nonumber\\
\Phi_N(t) &=& \Phi_c - \left[\frac{t_c - t}{5(1 + z){\cal
M}}\right]^{5/8}\;,\qquad f \equiv \frac{1}{\pi}\frac{d\Phi_N}{dt}\;.
\label{eq:NewtQuad}
\end{eqnarray}
Here $\Phi_N$ is the lowest-order contribution to the orbital phase,
$f$ is the GW frequency, and ${\cal M} = m_1^{3/5} m_2^{3/5}/(m_1 +
m_2)^{1/5}$ is the binary's ``chirp mass,'' which sets the rate at
which $f$ changes.  We use units with $G = c = 1$; handy conversion
factors are $M_\odot \equiv GM_\odot/c^2 = 1.47\,{\rm km}$, and
$M_\odot \equiv GM_\odot/c^3 = 4.92 \times 10^{-6}\,{\rm seconds}$.
The angle $\iota$ describes the inclination of the binary's orbital
plane to our line-of-sight: $\cos\iota = \mathbf{\hat
L}\cdot\mathbf{\hat n}$, where $\mathbf{\hat L}$ is the unit vector
normal to the binary's orbital plane, and $\mathbf{\hat n}$ is the
unit vector along the line-of-sight to the binary.  The parameters
$t_c$ and $\Phi_c$ are the time and orbital phase when 
$f$ diverges in this model. We expect finite size effects to impact
the waveform before this divergence is reached.

A given detector measures a linear combination of the polarizations:
\begin{equation}
h_{\rm meas} = F_+(\theta, \phi, \psi) h_+ + F_\times(\theta, \phi,
\psi) h_\times\;,
\label{eq:hmeas}
\end{equation}
where $\theta$ and $\phi$ describe the binary's position on the sky,
and the ``polarization angle'' $\psi$ sets the inclination of the
components of $\mathbf{\hat L}$ orthogonal to $\mathbf{\hat n}$.  The
angles $\iota$ and $\psi$ fully specify the orientation vector
$\mathbf{\hat L}$.  For a particular detector geometry, the antenna
functions $F_+$ and $F_\times$ can be found in~\cite{300yrs}.  In
Sec.\ \ref{sec:gwmeasure} we give a general form for the gravitational
waveform without appealing to a specific detector, following the
analysis of \citealt{cf94} (hereafter abbreviated CF94).

Several features of Eqs.\ (\ref{eq:NewtQuad}) and (\ref{eq:hmeas}) are
worth commenting upon.  First, note that the phase depends on the {\it
redshifted}\/ chirp mass.  Measuring phase thus determines the
combination $(1 + z){\cal M}$~\citep{fc93}, not ${\cal M}$ or $z$
independently.  To understand this, note that ${\cal M}$ controls how
fast the frequency evolves: using Eq.\ (\ref{eq:NewtQuad}), we find
$\dot f \propto f^{11/3}{\cal M}^{5/3}$.  The chirp mass enters the
system's dynamics as a timescale $\tau_c = G{\cal M}/c^3$.  For a
source at cosmological distance, this timescale is redshifted; the
chirp mass we infer is likewise redshifted.  Redshift and chirp mass
are inextricably degenerate.  This remains true even when higher order
effects (see, e.g., \citealt{blanchet06}) are taken into account:
parameters describing a binary impact its dynamics as timescales which
undergo cosmological redshift, so we infer redshifted values for those
parameters. {\it GW observations on their own cannot directly
determine a source's redshift.}

Next, note that the amplitude depends on $(1 + z){\cal M}$, the angles
$(\theta, \phi, \iota, \psi)$, and the luminosity distance $D_L$.
Measuring the amplitude thus measures a combination of these
parameters.  By measuring the phase, we measure the redshifted chirp
mass sufficiently well that $(1 + z){\cal M}$ essentially decouples
from the amplitude.  More concretely, matched filtering the data with
waveform templates should allow us to determine the phase with
fractional accuracy $\delta\Phi/\Phi \sim 1/[(\mbox{signal-to-noise})
\times (\mbox{number of measured cycles})]$; $(1 + z){\cal M}$ should
be measured with similar fractional accuracy.  NS-NS binaries will
radiate roughly $10^4$ cycles in the band of advanced LIGO, and NS-BH
binaries roughly $10^3$ cycles, so the accuracy with which phase and
redshifted chirp mass can be determined should be exquisite
(\citealt{fc93}, CF94).

Although $(1 + z){\cal M}$ decouples from the amplitude, the distance,
position, and orientation angles remain highly coupled.  To determine
source distance we must break the degeneracy that the amplitude's
functional form sets on these parameters.  One way to break these
degeneracies is to measure the waves with multiple detectors.  Studies
{\citep{sylvestre, cavalieretal, blairetal, fairhurst09, wenchen10}}
have shown that doing so allows us to determine source position to
within a few degrees in the best cases, giving some information about
the source's distance and inclination.

Perhaps the best way to break some of these degeneracies is to measure
the event electromagnetically.  An EM signature will pin down the
event's position far more accurately than GWs alone.  The position
angles then decouple, much as the redshifted chirp mass decoupled.
Using multiple detectors, we can then determine the source's
orientation and its distance.  This gives us a direct,
calibration-free measure of the distance to a cosmic event.  The EM
signature may also provide us with the event's redshift, directly
putting a point on the Hubble diagram. In addition, if
modeling or observation give us evidence for beaming of the SHB
emission, this could strongly constrain the source inclination.

\subsection{This work and previous analysis}
\label{sec:dalaletal}

Our goal is to assess how well we can determine the luminosity
distance $D_L$ to SHBs under the assumption that they are associated
with inspiral GWs.  We consider both NS-NS and NS-BH mergers as
generators of SHBs, and consider several plausible advanced detector
networks: the current LIGO/Virgo network, upgraded to advanced
sensitivity; LIGO/Virgo plus the proposed Australian AIGO; LIGO/Virgo
plus the proposed Japanese LCGT; and LIGO/Virgo plus AIGO plus LCGT.

The engine of our analysis is a probability function that describes
how inferred source parameters $\boldsymbol{\theta}$ should be
distributed following GW measurement.  (Components $\theta^a$ of the
vector $\boldsymbol{\theta}$ are physical parameters such as a
binary's masses, distance, sky position angles, etc.; our particular
focus is on $D_L$.)  Consider one detector which measures a datastream
$s(t)$, containing noise $n(t)$ and a GW signal
$h(t,{\boldsymbol{\hat\theta}})$, where $\boldsymbol{\hat\theta}$
describes the source's ``true'' parameters.  In the language of
\cite{finn92}, we assume ``detection'' has already occurred; our goal
in this paper is to focus on the complementary problem of
``measurement.''

As shown by {\cite{finn92}}, given a model for our signal
$h(t,\boldsymbol{\theta})$, and assuming that the noise statistics are
Gaussian, the probability that the parameters $\boldsymbol{\theta}$
describe the data $s$ is
\begin{equation}
p(\boldsymbol{\theta} | s) =
p_0(\boldsymbol{\theta})\exp\left[-\left(( h(\boldsymbol{\theta}) - s)
|( h(\boldsymbol{\theta}) - s )\right)/2\right]\;.
\label{eq:likelihood}
\end{equation}
The inner product $(a|b)$ describes the noise weighted
cross-correlation of $a(t)$ with $b(t)$, and is defined precisely
below.  The distribution $p_0(\boldsymbol{\theta})$ is a {\it prior
probability distribution}; it encapsulates what we know about our
signal prior to measurement.  We define $\boldsymbol{\tilde\theta}$ to
be the parameters that maximize Eq.\ (\ref{eq:likelihood}).

DHHJ06 did a first pass on the analysis we describe here.  They
expanded the exponential to second order in the variables
$(\boldsymbol{\theta} - \boldsymbol{\hat\theta})$; we will henceforth
refer to this as the ``Gaussian'' approximation (cf.\
\citealt{finn92}):
\begin{eqnarray}
\label{eq:gaussian}
& &\exp\left[-\left( h(\boldsymbol{\theta}) - s | h(\boldsymbol{\theta})
  - s\right)/2\right] \simeq
\nonumber\\
& &\qquad\qquad\qquad
\exp\left[-\frac{1}{2}\left(\frac{\partial h}{\partial\theta^a} \Biggl|
  \frac{\partial h}{\partial\theta^b}\right)\delta\theta^a
  \delta\theta^b\right]\;,
\end{eqnarray}
where $\delta\theta^a = \theta^a - \hat\theta^a$.  In this limit,
$\boldsymbol{\tilde\theta} = \boldsymbol{\hat\theta}$ (at least for
uniform priors).  The matrix
\begin{equation}
\Gamma_{ab} \equiv \left(\frac{\partial h}{\partial\theta^a} \Biggl|
  \frac{\partial h}{\partial\theta^b}\right)
\label{eq:Fisherdef}
\end{equation}
is the {\it Fisher information matrix}.  Its inverse $\Sigma^{ab}$ is
the covariance matrix.  Diagonal entries $\Sigma^{aa}$ are the
variance of parameter $\theta^a$; off-diagonal entries describe
correlations.

The Gaussian approximation to Eq.\ (\ref{eq:likelihood}) is known to
be accurate when the signal-to-noise ratio (SNR) is large.  However,
it is not clear what ``large'' really means {\citep{vallis}}.  Given
current binary coalescence rate estimates, it is expected that most
events will come from $D_L \sim \mbox{a few} \times 100\,{\rm Mpc}$.
In such cases, we can expect an advanced detector SNR $\sim 10$.  It
is likely that this value is not high enough for the ``large SNR''
approximation to be appropriate.

In this analysis we avoid the Gaussian approximation.  We instead
use Markov-Chain Monte-Carlo (MCMC) techniques (in particular, the
Metropolis-Hastings algorithm) to explore our parameter distributions.
A brief description of this technique is given in Sec.\
\ref{sec:estimate}, and described in detail in \cite{lewis02}.  We
find that the Gaussian approximation to Eq.\ (\ref{eq:likelihood}) is
indeed failing in its estimate of extrinsic parameters (though it
appears to do well for intrinsic parameters such as mass).

\subsection{Organization of this paper}

We begin in Sec.\ {\ref{sec:inspiralwaves}} by summarizing how GWs
encode the distance to a coalescing binary.  We first describe the
post-Newtonian (PN) gravitational waveform we use in Sec.\
{\ref{sec:gwform}}, and then describe how that wave interacts with a
network of detectors in Sec.\ {\ref{sec:gwmeasure}}.  Our discussion
of the network-wave interaction is heavily based on the notation and
formalism used in Sec.\ 4 of CF94, as well as the analysis of
\cite{abcf01}.  Section {\ref{sec:gwmeasure}} is sufficiently dense
that we summarize its major points in Sec.\
{\ref{sec:gwmeasure_summary}} before concluding, in Sec.\
{\ref{sec:detectors}}, with a description of the GW detectors which we
include in our analysis.

We outline parameter estimation in Sec.\ \ref{sec:estimate}.  In Sec.\
{\ref{sec:formaloverview}} we describe in more detail how to construct
the probability distributions describing parameter measurement.  We
then give, in Sec.\ \ref{sec:selectionandpriors}, a brief description
of our selection procedure based on SNR detection thresholds.  This
procedure sets physically motivated priors for some of our parameters.
The Markov-Chain Monte-Carlo technique we use to explore this function
is described in Sec.\ {\ref{sec:mhmc}}.  How to appropriately average
this distribution to give ``noise averaged'' results and to compare
with previous literature is discussed in Sec.\
{\ref{sec:averagedPDF}}.

In Sec.\ {\ref{sec:valid}} we discuss the validation of our code.  We
begin by attempting to reproduce some of the key results on distance
measurement presented in CF94.  Because of the rather different
techniques used by Cutler \& Flanagan, we do not expect exact
agreement.  It is reassuring to find, nonetheless, that we can
reconstruct with good accuracy all of the major features of their
analysis.  We then examine how these results change as we vary the
amplitude (moving a fiducial test binary to smaller and larger
distances), as we vary the number of detectors in our network, and as
we vary the source's inclination.

Our main results are given in Sec.\ {\ref{sec:main_results}}.  We
consider several different plausible detector networks and examine
measurement errors for two ``fiducial'' binary systems, comprising
either two neutron stars (NS-NS) with physical masses of $m_1 = m_2 =
1.4\,M_{\odot}$, or a neutron star and black hole (NS-BH) system with
physical masses $m_1 = 1.4\,M_{\odot}$ and $m_2 = 10\,M_{\odot}$.
Assuming a constant comoving cosmological density, we distribute
potential GW-SHB events on the sky, and select from this distribution
using a detection threshold criterion set for the entire GW detector
network.  We summarize some implications of our results in Sec.\
\ref{sec:summary}.  A more in-depth discussion of these implications,
particularly with regard to what they imply for cosmological
measurements, will be presented in a companion paper.

Throughout this paper, we use units with $G = c = 1$.  We define
the shorthand $m_z = (1 + z)m$ for any mass parameter $m$.

\section{Measuring gravitational waves from inspiraling binaries}
\label{sec:inspiralwaves}

In this section we review the GW description we use, the formalism
describing how these waves interact with a network of detectors, and
the properties of the detectors.

\subsection{GWs from inspiraling binaries}
\label{sec:gwform}

The inspiral and merger of a compact binary's members can be divided
into three consecutive phases.  The first and longest is a gradual
adiabatic {\it inspiral}, when the members slowly spiral together due
to the radiative loss of orbital energy and angular momentum.
Post-Newtonian (PN) techniques (an expansion in gravitational
potential $M/r$, or equivalently for bound systems, orbital speed
$v^2$) allow a binary's evolution and its emitted GWs to be modeled
analytically to high order; see \cite{blanchet06} for a review.  When
the bodies come close together, the PN expansion is no longer valid,
and direct numerical calculation is required.  Recent breakthroughs in
numerical relativity now make it possible to fully model the
strong-field, dynamical {\it merger} of two bodies into one; see
\cite{pretorius05}, \cite{su06}, and \cite{etienne08} for discussion.
If the end state is a single black hole, the final waves from the
system should be described by a {\it ringdown} as the black hole
settles down to the Kerr solution.

In this work we are concerned solely with the inspiral, and will
accordingly use the PN waveform to describe our waves.  In particular,
we use the so-called ``restricted'' PN waveform; following CF94, the
inspiral waveform may be written schematically
\begin{equation}
h(t) = \mathrm{Re}\left(\sum_{x,m}
h^x_m(t)e^{im\Phi_{\mathrm{orb}}(t)}\right) \, .
\label{eq:hPN}
\end{equation}
Here $x$ indicates PN order [$h^x$ is computed to $O(v^{2x})$ in
orbital speed], $m$ denotes harmonic order (e.g., $m = 2$ is
quadrupole), and $\Phi_{\rm orb}(t) = \int^t \Omega(t') dt'$ is
orbital phase [with $\Omega(t)$ the orbital angular frequency].  The
``restricted'' waveform neglects all PN amplitude terms beyond the
leading one, and considers only the dominant $m = 2$ contribution to
the phase.  The phase is computed to high PN order.

Let the unit vector $\hat{\bf n}$ point to a binary on the sky (so
that the waves propagate to us along $-\hat{\bf n}$), and let the unit
vector $\hat{\bf L}$ denote the normal along the binary's orbital
angular momentum.  The waveform is fully described by the two
polarizations:
\begin{eqnarray}
h_+(t) & = & \frac{2{\mathcal M}_z}{D_L}
\left[\pi{\cal M}_z f(t)\right]^{2/3}
[1+(\mathbf{\hat{L}}\cdot \mathbf{\hat{n}})^2]\cos[\Phi(t)]\;,
\nonumber\\
&\equiv& \frac{4{\mathcal M}_z}{D_L}
\left[\pi{\cal M}_z f(t)\right]^{2/3}
{\cal A}_+(\hat{\bf n},\hat{\bf L})\cos[\Phi(t)]\;;
\label{eq:hplus}\\
h_{\times}(t) & = & - \frac{4{\mathcal M}_z}{D_L}
[\pi{\cal M}_z f(t)]^{2/3}
(\mathbf{\hat{L}}\cdot \mathbf{\hat{n}}) \sin[\Phi(t)]\;,
\nonumber\\
&\equiv& \frac{4{\mathcal M}_z}{D_L}
\left[\pi{\cal M}_z f(t)\right]^{2/3}
{\cal A}_\times(\hat{\bf n},\hat{\bf L})\sin[\Phi(t)]\;.
\label{eq:hcross}
\end{eqnarray}
Equations (\ref{eq:hplus}) and (\ref{eq:hcross}) are nearly identical
to those given in Eq.\ (\ref{eq:NewtQuad}); only the phase $\Phi(t)$
is different, as described below.  ${\cal M}_z$ is the binary's
redshifted chirp mass, $D_L$ is its luminosity distance, and we have
written the inclination angle $\cos\iota$ using the vectors $\hat{\bf
n}$ and $\hat{\bf L}$.  The functions ${\cal A}_{+,\times}$ compactly
gather all dependence on sky position and orientation.  In Sec.\
{\ref{sec:gwmeasure}} we discuss how these polarizations interact
with our detectors.

In these forms of $h_+$ and $h_\times$, the phase is computed to
2nd-post-Newtonian (2PN) order \citep{bdiww95}:
\begin{eqnarray}
\Phi(t) &=& 2\pi \int f(t')\,dt' = 2\pi \int \frac{f}{df/dt}df\;,
\\
\frac{df}{dt} &=& \frac{96}{5}\pi^{8/3}
\mathcal{M}_z^{5/3}f^{11/3}\left[1 -
\left(\frac{743}{336} + \frac{11}{4}\eta\right)(\pi M_z f)^{2/3}
\right.
\nonumber\\
& &\quad\left. +
4\pi(\pi M_z f) \right. \nonumber \\
& &\quad \left. +
\left(\frac{34103}{18144} + \frac{13661}{2016}\eta +
\frac{59}{18}\eta^2 \right)(\pi M_z f)^{4/3}\right] \;.
\label{eq:freqchirp}
\end{eqnarray}
Higher order results for $df/dt$ are now known (\citealt{bij02,
bfij02, blanchet04}), but 2PN order will be adequate for our purposes.
Since distance measurements depend on accurate amplitude
determination, we do not need a highly refined model of the wave's
phase.  The rate of sweep is dominantly determined by the chirp mass,
but there is an important correction due to $\eta = \mu /M = m_1
m_2/(m_1 + m_2)^2$, the reduced mass ratio.  Note that $\eta$ is not
redshifted; both $\mu$ and $M$ (the reduced mass and total mass,
respectively) acquire $(1 + z)$ corrections, so their ratio is the
same at all $z$.  Accurate measurement of the frequency sweep can thus
determine both ${\cal M}_z$ and $\eta$ (or ${\cal M}_z$ and $\mu_z$).

We will find it useful to work in the frequency domain, using the
Fourier transform ${\tilde h}(f)$ rather than $h(t)$:
\begin{equation}
{\tilde h}(f) \equiv \int_{-\infty}^{\infty}\, e^{2\pi i f t}h(t)\, dt\;.
\label{eq:fourierT}
\end{equation}
An approximate result for ${\tilde h}(f)$ can be found using
stationary phase \citep{fc93}, which describes the Fourier transform
when $f$ changes slowly:
\begin{eqnarray}
\tilde{h}_+(f) &=& \sqrt{\frac{5}{96}}\frac{\pi^{-2/3} {\mathcal
M}_z^{5/6}}{D_L}{\cal A}_+ f^{-7/6} e^{i\Psi(f)} \, ,
\label{eq:freqdomainhp}
\\
\tilde{h}_\times(f) &=& \sqrt{\frac{5}{96}}\frac{\pi^{-2/3} {\mathcal
M}_z^{5/6}}{D_L}{\cal A}_\times f^{-7/6} e^{i\Psi(f) - i\pi/2} \, .
\label{eq:freqdomainhc}
\end{eqnarray}
``Slowly'' means that $f$ does not change very much over a single wave
period $1/f$, so that $(df/dt)/f \ll f$.  The validity of this
approximation for the waveforms we consider, at least until the last
moments before merger, has been demonstrated in previous work
{\citep{droz99}}.  The phase function $\Psi(f)$ in Eqs.\
(\ref{eq:freqdomainhp}) and (\ref{eq:freqdomainhc}) is given by
\begin{eqnarray}
\Psi(f) &=& 2\pi f t_c - \Phi_c - \frac{\pi}{4} + \frac{3}{128}(\pi
{\mathcal M} f)^{-5/3}
\times\nonumber\\
& &\left[1 + \frac{20}{9}\left(\frac{743}{336} +
\frac{11}{4}\eta\right)(\pi M_z f)^{2/3} \right.
\nonumber\\
& &\left. -16\pi( \pi M_z f)\right.
\nonumber\\
& &\left.+ 10\left(\frac{3058673}{1016064} +
\frac{5429}{1008}\eta + \frac{617}{144}\eta^2 \right)(\pi
M_z f)^{4/3}\right] \, .
\nonumber\\
\label{eq:PNpsi}
\end{eqnarray}
As in Eq.\ (\ref{eq:NewtQuad}), $t_c$ is called the ``time of
coalescence'' and defines the time at which $f$ diverges within the PN
framework; $\Phi_c$ is similarly the ``phase at coalescence.''  We
assume an abrupt and unphysical transition between inspiral and merger
at the innermost stable circular orbit (ISCO), $f_{\rm ISCO}=(6
\sqrt{6} \pi M_z)^{-1}$.  For NS-NS, $f_{\rm ISCO}$ occurs at high
frequencies where detectors have poor sensitivity.  As such, we are
confident that this abrupt transition has little impact on our
results.  For NS-BH, $f_{\rm ISCO}$ is likely to be in a band with
good sensitivity, and better modeling of this transition will be
important.

In this analysis we neglect effects which depend on spin.  In general
relativity, spin drives precessions which can ``color'' the waveform
in important ways, and which can have important observational effects
(see, e.g., \citealt{v04}, \citealt{lh06}, \citealt{vandersluys08}).
These effects are important when the dimensionless spin parameter, $a
\equiv c|{\bf S}|/GM^2$, is fairly large.  Neutron stars are unlikely
to spin fast enough to drive interesting precession during the time
that they are in the band of GW detectors.  To show this, write the
moment of inertia of a neutron star as
\begin{equation}
I_{\rm NS} = \frac{2}{5}\kappa M_{\rm NS} R_{\rm NS}^2\;,
\end{equation}
where $M_{\rm NS}$ and $R_{\rm NS}$ are the star's mass and radius,
and the parameter $\kappa$ describes the extent to which its mass is
centrally condensed (compared to a uniform sphere).  Detailed
calculations with different equations of state indicate $\kappa \sim
0.7$--$1$ [cf.\ \cite{cook94}, especially the slowly rotating
configurations in their Tables 12, 15, 18, and 21].  For a neutron
star whose spin period is $P_{\rm NS}$, the Kerr parameter is given by
\begin{eqnarray}
a_{\rm NS} &=& \frac{c}{G}\frac{I_{\rm NS}}{M_{\rm NS}^2}
\frac{2\pi}{P_{\rm NS}}
\nonumber\\
&\simeq& 0.06 \kappa \left(\frac{R_{\rm
NS}}{\rm{12\,km}}\right)^2\left(\frac{1.4\,M_\odot}{M_{\rm NS}}\right)
\left(\frac{10\,{\rm msec}}{P_{\rm NS}}\right)\;.
\end{eqnarray}
As long as the neutron star spin period is longer than $\sim10$ msec,
$a_{\rm NS}$ is small enough that spin effects can be neglected in our
analysis.  We {\it should}\/ include spin in our models of BH-NS
binaries; we leave this to a later analysis.  Van der Sluys et al.\
(2008) included black hole spin effects in an analysis which did not
assume known source position. They found that spin-induced modulations
could help GW detectors to localize a source.  This and companion
works (\citealt{raymond09}, \citealt{vandersluys09}) suggest that, if
position is known, spin modulations could improve our ability to
measure source inclination and distance.

Our GWs depend on nine parameters: two masses ${\cal M}_z$ and
$\mu_z$, two sky position angles (which set $\hat{\bf n}$), two
orientation angles (which set $\hat{\bf L}$), time at coalescence
$t_c$, phase at coalescence $\Phi_c$, and luminosity distance $D_L$.
When sky position is known, the parameter set is reduced to seven: $\{
{\cal M}_z, \mu_z, D_L, t_c, \cos \iota, \psi, \Phi_c \}$.

\subsection{Measurement of GWs by a detector network}
\label{sec:gwmeasure}

We now examine how the waves described in Sec.\ {\ref{sec:gwform}}
interact with a network of detectors.  We begin by introducing a
geometric convention, which follows that introduced in CF94 and in
\cite{abcf01}.  A source's sky position is given by a unit vector
$\hat{\bf n}$ (which points from the center of the Earth to the
binary), and its orientation is given by a unit vector $\hat{\bf L}$
(which points along the binary's orbital angular momentum).  We
construct a pair of axes which describe the binary's orbital plane:
\begin{equation}
\hat{\bf X} = \frac{\hat{\bf n}\times\hat{\bf L}}{|\hat{\bf
n}\times\hat{\bf L}|}\;,\quad
\hat{\bf Y} = -\frac{\hat{\bf n}\times\hat{\bf X}}{|\hat{\bf
n}\times\hat{\bf X}|}\;.
\label{eq:XYvectors}
\end{equation}
With these axes, we define the {\it polarization basis tensors}
\begin{eqnarray}
{\bf e}^+ &=& \hat{\bf X} \otimes \hat{\bf X} - \hat{\bf Y} \otimes
\hat{\bf Y}\;,
\label{eq:plusbasis}
\\
{\bf e}^\times &=& \hat{\bf X} \otimes \hat{\bf Y} + \hat{\bf Y}
\otimes \hat{\bf X}\;.
\label{eq:timesbasis}
\end{eqnarray}
The transverse-traceless metric perturbation describing our source's
GWs is then
\begin{equation}
h_{ij} = h_+ e^+_{ij} + h_\times e^\times_{ij}\;.
\label{eq:wavetensor}
\end{equation}

We next characterize the GW detectors.  Each detector is an $L$-shaped
interferometer whose arms define two-thirds of an orthonormal triple.
Denote by $\hat{\bf x}_a$ and $\hat{\bf y}_a$ the unit vectors along
the arms of the $a$-th detector in our network; we call these the $x$-
and $y$-arms.  (The vector $\hat{\bf z}_a = \hat{\bf x}_a \times
\hat{\bf y}_a$ points radially from the center of the Earth to the
detector's vertex.)  These vectors define the {\it response tensor}\/
for detector $a$:
\begin{equation}
D^{ij}_a = \frac{1}{2}\left[(\hat{\bf x}_a)^i (\hat{\bf x}_a)^j -
(\hat{\bf y}_a)^i (\hat{\bf y}_a)^j\right]\;.
\label{eq:detresponse}
\end{equation}
The response of detector $a$ to a GW is given by
\begin{eqnarray}
h_a &=& D^{ij}_a h_{ij}
\nonumber\\
&\equiv& e^{- 2 \pi i ({\bf n}\cdot{\bf r}_a) f} (F_{a,+}h_+ +
F_{a,\times}h_\times)\;,
\label{eq:measuredwave}
\end{eqnarray}
where $\bf{r}_a$ is the position of the detector $a$ and the factor
$({\bf n}\cdot {\bf r}_a)$ measures the time of flight between it and
the coordinate origin.  The second form of Eq.\
(\ref{eq:measuredwave}) shows how the antenna functions introduced in
Eq.\ (\ref{eq:hmeas}) are built from the wave tensor and the response
tensor.

Our discussion has so far been frame-independent, in that we have
defined all vectors and tensors without reference to coordinates.  We
now introduce a coordinate system for our detectors following
\cite{abcf01} [who in turn use the WGS-84 Earth model
{\citep{althouse_etal}}].  The Earth is taken to be an oblate
ellipsoid with semi-major axis $a = 6.378137 \times 10^6$ meters, and
semi-minor axis $b = 6.356752314 \times 10^6$ meters.  Our coordinates
are fixed relative to the center of the Earth.  The $x$-axis (which
points along ${\bf i}$) pierces the Earth at latitude $0^\circ$ North,
longitude $0^\circ$ East (normal to the equator at the prime
meridian); the $y$-axis (along ${\bf j}$) pierces the Earth at
$0^\circ$ North, $90^\circ$ East (normal to the equator in the Indian
ocean somewhat west of Indonesia); and the $z$-axis (along ${\bf k}$)
pierces the Earth at $90^\circ$ North (the North geographic pole).

A GW source at $(\theta,\phi)$ on the celestial sphere has sky
position vector $\hat{\bf n}$:
\begin{equation}
\hat{\bf n} = \sin\theta\cos\phi{\bf i} + \sin\theta\sin\phi{\bf j} +
\cos\theta{\bf k}\;.
\end{equation}
The {\it polarization angle}, $\psi$, is the angle (measured clockwise
about $\hat{\bf n}$) from the orbit's line of nodes to the source's
$\hat{\bf X}$-axis.  In terms of these angles, the vectors $\hat{\bf
X}$ and $\hat{\bf Y}$ are given by {\citep{abcf01}}
\begin{eqnarray}
\hat{\bf X} &=& (\sin\phi\cos\psi - \sin\psi\cos\phi\cos\theta) {\bf
i}\nonumber\\
& & - (\cos\phi\cos\psi + \sin\psi\sin\phi\cos\theta){\bf j}
+\sin\psi\sin\theta {\bf k}\;,
\nonumber\\
\label{eq:Xvector2}\\
\hat{\bf Y} &=& (-\sin\phi\sin\psi - \cos\psi\cos\phi\cos\theta) {\bf
i}\nonumber\\
& & + (\cos\phi\sin\psi - \cos\psi\sin\phi\cos\theta){\bf j} +
\cos\psi\sin\theta {\bf k}\;.
\nonumber\\
\label{eq:Yvector2}
\end{eqnarray}
The angle $\phi$ is related to right ascension $\alpha$ by $\alpha =
\phi + {\rm GMST}$ (where GMST is the Greenwich mean sidereal time at
which the signal arrives), and $\theta$ is related to declination
$\delta$ by $\delta = \pi/2 - \theta$ (cf.\ \citealt{abcf01}, Appendix
B).  Combining Eqs.\ (\ref{eq:Xvector2}) and (\ref{eq:Yvector2}) with
Eqs.\ (\ref{eq:plusbasis})--(\ref{eq:wavetensor}) allows us to write
$h_{ij}$ for a source in coordinates adapted to this problem.

We now similarly describe our detectors using convenient coordinates.
Detector $a$ is at East longitude $\lambda_a$ and North latitude
$\varphi_a$ (not to be confused with sky position angle $\phi$).  The
unit vectors pointing East, North, and Up for this detector are
\begin{eqnarray}
{\bf e}^{\rm E}_a &=& -\sin\lambda_a{\bf i} + \cos\lambda_a{\bf j}\;,
\label{eq:eastunitvec}
\\
{\bf e}^{\rm N}_a &=& -\sin\varphi_a\cos\lambda_a{\bf i} -
\sin\varphi_a\sin\lambda_a{\bf j} + \cos\varphi_a{\bf k}\;,
\label{eq:northunitvec}
\\
{\bf e}^{\rm U}_a &=& \cos\varphi_a\cos\lambda_a{\bf i} +
\cos\varphi_a\sin\lambda_a{\bf j} - \cos\varphi_a{\bf k}\;.
\label{eq:upunitvec}
\end{eqnarray}
The $x$-arm of detector $a$ is oriented at angle $\Upsilon_a$ North of
East, while its $y$-arm is at angle $\Upsilon_a + \pi/2$.  Thanks to
the Earth's oblateness, the $x$- and $y$-arms are tilted at angles
$\omega^{x,y}_a$ to the vertical.  The unit vectors $\hat{\bf x}_a$,
$\hat{\bf y}_a$ can thus be written
\begin{eqnarray}
\hat{\bf x}_a &=& \cos\omega^x_a \cos\Upsilon_a{\bf e}^{\rm E}_a +
\cos\omega^x_a\sin\Upsilon_a{\bf e}^{\rm N}_a + \sin\omega^x_a{\bf
e}^{\rm U}\;,
\nonumber\\
\label{eq:detectorxhat}
\\
\hat{\bf y}_a &=& -\cos\omega^y_a\sin\Upsilon_a{\bf e}^{\rm E}_a +
\cos\omega^y_a\cos\Upsilon_a{\bf e}^{\rm N}_a + \sin\omega^y_a{\bf
e}^{\rm U}\;.
\nonumber\\
\label{eq:detectoryhat}
\end{eqnarray}
Combining Eqs.\ (\ref{eq:detectorxhat}) and (\ref{eq:detectoryhat})
with Eq.\ (\ref{eq:detresponse}) allows us to write the response
tensor for each detector in our network.

\subsection{Summary of the preceding section}
\label{sec:gwmeasure_summary}

Section {\ref{sec:gwmeasure}} is sufficiently dense that a brief
summary may clarify its key features, particularly with respect to the
quantities we hope to measure.  From Eq.\ (\ref{eq:measuredwave}), we
find that each detector in our network measures a weighted sum of the
two GW polarizations $h_+$ and $h_\times$.  Following \cite{cutler98},
we can rewrite the waveform detector $a$ measures as
\begin{equation}
h_a = \frac{4{\cal M}_z}{D_L}{\cal A}_p\left[\pi {\cal M}_z
f(t)\right]^{2/3} \cos\left[\Phi(t) + \Phi_p\right]\;,
\label{eq:measuredwave2}
\end{equation}
where we have introduced detector $a$'s ``polarization amplitude''
\begin{equation}
{\cal A}_p = \sqrt{ \left(F_{a,+}{\cal A}_+\right)^2 +
\left(F_{a,\times}{\cal A}_\times\right)^2}\;,
\label{eq:polamp}
\end{equation}
and its ``polarization phase''
\begin{equation}
\tan\Phi_p = \frac{F_{a,\times}{\cal A}_\times}{F_{a,+}{\cal A}_+}\;.
\label{eq:polphase}
\end{equation}
The intrinsic GW phase, $\Phi(t)$, is a strong function of the
redshifted chirp mass, ${\cal M}_z$, the redshifted reduced mass,
$\mu_z$, the time of coalescence, $t_c$, and the phase at coalescence,
$\Phi_c$.  Measuring the phase determines these four quantities,
typically with very good accuracy.

Consider for a moment measurements by a single detector.  The
polarization amplitude and phase depend on the binary's sky position,
$(\theta,\phi)$ or $\hat{\bf n}$, and orientation, $(\psi,\iota)$ or
$\hat{\bf L}$. [They also depend on detector position, $(\lambda_a,
\varphi_a)$, orientation, $\Upsilon_a$, and tilt, $(\omega^x_a,
\omega^y_a)$.  These angles are known and fixed, so we ignore them in
this discussion.]  If the angles $(\theta,\phi,\psi,\iota)$ are not
known, a single detector cannot separate them, nor can it separate the
distance $D_L$.

Multiple detectors can, at least in principle, separately determine
these parameters.  Each detector measures its own amplitude and
polarization phase.  Combining their outputs, we can fit to the
unknown angles and the distance.  Various works have analyzed how well
this can be done assuming that the position and orientation are
completely unknown (\citealt{sylvestre, cavalieretal, blairetal}).
Van der Sluys et al.\ (2008) performed such an analysis for
measurements of NS-BH binaries, including the effect of orbital
precession induced by the black hole.  This precession effectively
make the angles $\iota$ and $\psi$ time dependent, also breaking the
degeneracy among these angles and $D_L$.

In what follows, we assume that an electromagnetic identification pins
down the angles $(\theta,\phi)$, so that they do not need to be
determined from the GW data.  We then face the substantially less
challenging problem of determining $\psi$, $\iota$, and $D_L$.  We
will also examine the impact of a constraint on the inclination,
$\iota$.  Long bursts are believed to be strongly collimated, emitting
into jets with opening angles of just a few degrees.  Less is known
about the collimation of SHBs, but it is plausible that their emission
may be primarily along a preferred axis (presumably the progenitor
binary's orbital angular momentum axis).

\subsection{GW detectors used in our analysis}
\label{sec:detectors}

Here we briefly summarize the properties of the GW detectors that we
consider.

\begin{figure}
\centering 
\includegraphics[angle=90,width=0.98\columnwidth]{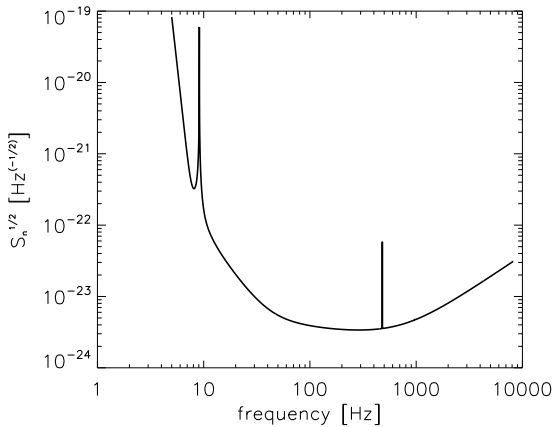}
\caption{Anticipated noise spectrum for Advanced LIGO
({\citealt{ligo_noise}}; cf.\ their Fig.\ 3).  Our calculations assume
no astrophysically interesting sensitivity below a low frequency
cut-off of 10 Hz.  The features at $f \simeq 10$ Hz and a few hundred
Hz are resonant modes of the mirror suspensions driven by thermal
noise.}
\label{fig:aligonoise}
\end{figure}

\noindent
{\it LIGO}: The Laser Interferometer Gravitational-wave Observatory
consists of two 4 kilometer interferometers located in Hanford,
Washington (US) and Livingston, Louisiana (US).  These instruments
have achieved their initial sensitivity goals.  An upgrade to
``advanced'' configuration is expected to be completed around 2014,
with tuning for best sensitivity to be undertaken in the years
following\footnote{http://www.ligo.caltech.edu/advLIGO/scripts/summary.shtml}.
We show the anticipated noise limits from fundamental noise sources in Fig.\
{\ref{fig:aligonoise}} for a broad-band tuning {\citep{ligo_noise}}.
This spectrum is expected to be dominated by quantum sensing noise
above a cut-off at $f < 10$ Hz, with a contribution from thermal noise
in the test mass coatings in the band from 30--200 Hz.

\noindent
{\it Virgo}: The Virgo detector \citep{acernese08} near Pisa, Italy has slightly shorter
arms than LIGO (3 kilometers), but should achieve similar advanced
sensitivity on roughly the same timescale as the LIGO
detectors\footnote{http://www.ego-gw.it/public/virgo/virgo.aspx}.  For
simplicity, we will take Virgo's sensitivity to be the same as LIGO's.

Our baseline detector network consists of the LIGO Hanford and
Livingston sites, and Virgo; these are instruments which are running
today, and will be upgraded over the next decade.  We also examine the
impact of adding two proposed interferometers to this network:

\noindent
{\it AIGO}: The Australian International Gravitational Observatory \citep{barriga10} is
a proposed multi-kilometer interferometer that would be located in
Gingin, Western Australia.  AIGO's proposed site in Western Australia
is particularly favorable due to low seismic and human activity.

\noindent
{\it LCGT}: The Large-scale Cryogenic Gravitational-wave Telescope \citep{kuroda10} is
a proposed multi-kilometer interferometer that would be located in the
Kamioka observatory, 1 kilometer underground.  This location takes
advantage of the fact that local ground motions tend to decay rapidly
as we move away from the Earth's surface.  They also plan to use
cryogenic cooling to reduce thermal noise.

As with Virgo, we will take the sensitivity of AIGO and LCGT to be the
same as LIGO for our analysis.  Table {\ref{tab:detectors}} gives the
location and orientation of these detectors, needed to compute
each detector's response function.  It's worth mentioning that more
advanced detectors are in the early planning stages.  Particularly
noteworthy is the European proposal for the ``Einstein Telescope,''
currently undergoing design studies.  It is being designed to study
binary coalescence to high redshift ($z \gtrsim 5$) {\citep{sathya09}}.

\begin{widetext}

\begin{deluxetable}{lccccc}
\tablewidth{17.5cm}
\tablecaption{GW detectors (positions and orientations).}
\tablehead{
\colhead{Detector} &
\colhead{East Long.\ $\lambda$} &
\colhead{North Lat.\ $\varphi$} &
\colhead{Orientation $\Upsilon$} &
\colhead{$x$-arm tilt $\omega^x$} &
\colhead{$y$-arm tilt $\omega^y$}
}

\startdata

LIGO-Han & $-119.4^\circ$ & $46.5^\circ$ & $126^\circ$ & $ (-6.20 \times 10^{-4})^\circ$ & $(1.25 \times 10^{-5})^\circ$ \\
LIGO-Liv & $-90.8^\circ$ & $30.6^\circ$ & $198^\circ$ & $ (-3.12 \times 10^{-4})^\circ$ & $(-6.11 \times 10^{-4} )^\circ$ \\
Virgo & $10.5^\circ$ & $43.6^\circ$ & $70^\circ$ & $0.0^\circ$ & $0.0^\circ$ \\
AIGO & $115.7^\circ$ & $-31.4^\circ$ & $0^\circ$ & $0.0^\circ$ & $0.0^\circ$ \\
LCGT & $137.3^\circ$ & $36.4^\circ$ & $25^\circ$ & $0.0^\circ$ & $0.0^\circ$ \\

\enddata

\label{tab:detectors}
\end{deluxetable}
\end{widetext}

\section{Estimation of binary parameters}
\label{sec:estimate}

\subsection{Overview of formalism}
\label{sec:formaloverview}

We now give a brief summary of the parameter estimation formalism we
use.  Further details can be found in \cite{finn92}, \cite{krolak93},
and CF94.

Assuming detection has occurred, the datastream of detector $a$,
$s_a(t)$, has two contributions: The true GW signal
$h_a(t;{\boldsymbol{\hat \theta}})$ (constructed by contracting the GW
tensor $h_{ij}$ with detector $a$'s response tensor $D^{ij}_a$; cf.\
Sec.\ {\ref{sec:gwmeasure}}), and a realization of detector noise
$n_a(t)$,
\begin{equation}
\label{eq:sig}
s_a(t) = h_a(t; {\boldsymbol{\hat \theta}}) + n_a(t)\;.
\end{equation}
The incident gravitational wave strain depends on (unknown) true
parameters ${\boldsymbol{\hat \theta}}$.  As in Sec.\
\ref{sec:dalaletal}, $\boldsymbol{\hat\theta}$ is a vector whose
components are binary parameters.  Below we use a vector ${\bf s}$
whose components $s_a$ are the datastreams of each detector.
Likewise, ${\bf h}$ and ${\bf n}$ are vectors whose components are the
GW and noise content of each detector.

We assume the noise to be stationary, zero mean, and Gaussian.  This
lets us categorize it using the spectral density as follows.  First,
define the noise correlation matrix:
\begin{eqnarray}
C_n(\tau)_{ab} &=& \langle n_a(t + \tau) n_b(t) \rangle -
\langle n_a(t + \tau) \rangle \, \langle n_b(t) \rangle
\nonumber\\
&=& \langle n_a(t + \tau) n_b(t) \rangle\;,
\label{eq:autocov}
\end{eqnarray}
where the angle brackets are ensemble averages over noise
realizations, and the zero mean assumption gives us the simplified
form on the second line.  For $a = b$, this is the auto-correlation of
detector $a$'s noise; otherwise, it describes the correlation between
detectors $a$ and $b$.  The (one-sided) power spectral density matrix
is the Fourier transform of this:
\begin{equation}
\label{eq:sn_def}
S_n(f)_{ab} = 2 \int_{-\infty}^{\infty} d \tau \, e^{2 \pi i f \tau}
C_n(\tau)_{ab}\;.
\end{equation}
This is defined for $f > 0$ only.  For $a = b$, it is the spectral
density of noise power in detector $a$; for $a \ne b$, it again
describes correlations between detectors.  From these definitions, one
can show that
\begin{equation}
\label{eq:noisestats}
\langle {\tilde n}_a(f) \, {\tilde n}_b(f^\prime)^* \rangle = {1 \over 2}
\delta(f - f^\prime) S_n(f)_{ab}.
\end{equation}
For Gaussian noise, this statistic completely characterizes our
detector noise.  No real detector is completely Gaussian, but by using
multiple, widely-separated detectors non-Gaussian events can be
rejected.  For this analysis, we assume the detectors' noises are
uncorrelated such that Eq.\ (\ref{eq:noisestats}) becomes
\begin{equation}
\label{eq:noisestatsunocrrelated}
\langle {\tilde n}_a(f) \, {\tilde n}_b(f^\prime)^* \rangle = {1 \over
2} \delta_{ab} \delta(f - f^\prime) S_n(f)_a.
\end{equation}
Finally, for simplicity we assume that $S_n(f)_a$ has the universal
shape $S_n(f)$ projected for advanced LIGO, shown in Fig.\
\ref{fig:aligonoise}.

Many of our assumptions are idealized (Gaussian noise; identical noise
spectra; no correlated noise between interferometers), and will
certainly not be achieved in practice.  These idealizations greatly
simplify our analysis, however, and are a useful baseline.  It would
be useful to revisit these assumptions and understand the quantitative
impact that they have on our analysis, but we do not expect a major
qualitative change in our conclusions.

The central quantity of interest in parameter estimation is the
posterior probability distribution function (PDF) for
${\boldsymbol{\theta}}$ given detector output {\bf s}, which is
defined as
\begin{equation}
\label{eq:postPDF}
p({\boldsymbol \theta} \, | \, {\bf s}) = {\cal N} \, p^{(0)}
({\boldsymbol{\theta}}) {\cal L}_{\rm TOT} ({\bf s} \, |
\,{\boldsymbol{\theta}}) \,.
\end{equation}
${\cal N}$ is a normalization constant,
$p^{(0)}({\boldsymbol{\theta}})$ is the PDF that represents the prior
probability that a measured GW is described by the parameters
$\boldsymbol{\theta}$, and ${\cal L}_{\rm TOT} (\bf{s} \, | \,
{\boldsymbol \theta} )$ is the total {\it likelihood function} (e.g.,
\citealt{mackay03}).  The likelihood function measures the relative
conditional probability of observing a particular dataset $\bf{s}$
given a measured signal ${\bf h}$ depending on some unknown set of
parameters $\boldsymbol{\theta}$ and given noise ${\bf n}$.  Because
we assume that the noise is independent and uncorrelated at each
detector site, we may take the total likelihood function to be the
product of the individual likelihoods at each detector:
\begin{equation}
\label{eq:totLike}
{\cal L}_{\rm TOT} ({\bf s} \, | \,{\boldsymbol \theta}) = \Pi_{a}
{\cal L}_a (s_a \, | \,{\boldsymbol \theta})\;,
\end{equation}
where ${\cal L}_a$, the likelihood for detector $a$, is given by
\citep{finn92}
\begin{equation}
\label{eq:Like}
{\cal L}_a \, (s \, | \,{\boldsymbol \theta}) = \, e^{ -
\big( h_a({\boldsymbol \theta}) - s_a \, \big| \, h_a({\boldsymbol
\theta}) - s_a \big)/2 } \, .
\end{equation}
The inner product $\left( \ldots | \ldots \right)$ on the vector space
of signals is defined as
\begin{equation}
(g|h) = 2 \int_0^{\infty} df \frac{\tilde{g}^*(f)\tilde{h}(f) +
\tilde{g}(f)\tilde{h}^*(f)}{S_n(f)} \, .
\label{eq:innerproduct}
\end{equation}
This definition means that the probability of the noise $n(t)$
taking some realization $n_0(t)$ is
\begin{equation}
\label{eq:noise_distribution}
p(n = n_0) \, \propto \, e^{- \left( n_0 | n_0 \right) / 2 }.
\end{equation}
\noindent 

For clarity, we distinguish between various definitions of SNR.  The
{\it true}\/ SNR at detector $a$, associated with a given instance of
noise for a measurement at a particular detector, is defined as (CF94)
\begin{eqnarray}
\left({S\over N}\right)_{a, {\rm true}} & = & { \left( h_a \, |
\, s_a \right) \over \sqrt{ \left( h_a \, | \, h_a\right) } }\;.
\label{eq:snr_true}
\end{eqnarray}
This is a random variable with Gaussian PDF of unit variance. For an
ensemble of realizations of the detector noise $n_a$, the {\it
average} SNR at detector $a$ is given by
\begin{equation}
\label{eq:snr_ave}
\left({S\over N}\right)_{a, {\rm ave}} = {{(h_a | h_a)}\over
{ {\rm rms}\ (h_a|n_a)}} = (h_a|h_a)^{1/2}.
\end{equation}
Consequently, we can define the combined {\it true} and {\it average}
SNRs of a coherent network of detectors:
\begin{eqnarray}
\left({S\over N}\right)_{{\rm true}} & = & \sqrt{\sum_a \left({S\over
N}\right)^2_{a, {\rm true}}}\ \ ,
\label{eq:snr_total}
\end{eqnarray}
and
\begin{eqnarray}
\left({S\over N}\right)_{{\rm ave}} & = & \sqrt{\sum_a \left({S\over
N}\right)^2_{a, {\rm ave}}}\ \ .
\label{eq:snr_tot_ave}
\end{eqnarray}

Estimating the parameter set ${\boldsymbol{\theta}}$ is often done
using a ``maximum likelihood'' method following either a Bayesian
(\citealt{loredo89}, \citealt{finn92}, CF94, \citealt{poisson95}) or
frequentist point of view (\citealt{krolak93}, CF94).  We do not
attempt to review these philosophies, and instead refer to Appendix A2
of CF94 for detailed discussion.  It is worth noting that, in the GW
literature, the ``maximum likelihood'' or ``maximum a posterior'' are
often interchangeably referred to as ``best-fit'' parameters.  The
maximum a posterior is the parameter set
$\boldsymbol{\tilde\theta}_{\rm MAP}$ which maximizes the full
posterior probability, Eq.\ (\ref{eq:postPDF}); likewise, the maximum
likelihood is the parameter set $\boldsymbol{\tilde\theta}_{\rm ML}$
which maximizes the likelihood function, Eq.\ (\ref{eq:totLike}).

Following the approach advocated by CF94, we introduce the Bayes
estimator ${\tilde \theta}_{\rm BAYES}^i({\bf s})$,
\begin{equation}
\label{eq:Bayes}
{\tilde \theta}_{\rm BAYES}^i({\bf s}) \equiv \int {\theta}^i\,
p(\boldsymbol{\theta} \, | \, {\bf s}) d\boldsymbol{\theta}\;.
\end{equation}        
The integral is performed over the whole parameter set
$\boldsymbol{\theta}$; $d\boldsymbol{\theta} = d\theta^1d\theta^2\dots
d\theta^n$. Similarly, we define the rms measurement errors
$\Sigma_{\rm BAYES}^{ij}$
\begin{equation}
\label{eq:sigma_bayes}
\Sigma_{\rm BAYES}^{ij} = \int ({\theta}^i - {\tilde \theta}^i_{\rm
BAYES}) \, ({\theta}^j - {\tilde \theta}^j_{\rm BAYES}) \,
p(\boldsymbol{ \theta} \, | \, {\bf s}) d\boldsymbol{\theta} .
\end{equation}
To understand the meaning of ${\tilde\theta}_{\rm BAYES}^i({\bf s})$,
consider a single detector which records an arbitrarily large ensemble
of signals.  This ensemble will contain a sub-ensemble in which the
various $s(t)$ are identical to one another.  Each member of the
sub-ensemble corresponds to GW signals with different true parameters
$\boldsymbol{\hat \theta}$, but have noise realizations $n(t)$ that
conspire to produce the same $s(t)$.  In this case, ${\tilde
\theta}_{\rm BAYES}^i({\bf s})$ is the expectation of $\theta^i$
averaged over the sub-ensemble.  The principle disadvantage of the
Bayes estimator is the computational cost to evaluate the
multi-dimensional integrals in Eqs.\ (\ref{eq:Bayes}) and
(\ref{eq:sigma_bayes}).

For large SNR it can be shown that the estimators
$\boldsymbol{\tilde\theta}_{\rm ML}$, $\boldsymbol{\tilde\theta}_{\rm
MAP}$, and $\boldsymbol{\tilde\theta}_{\rm BAYES}$ agree with one
another (CF94), and that Eq.\ (\ref{eq:postPDF}) is well-described by
a Gaussian form [cf.\ Eq.\ (\ref{eq:gaussian})].  However, as
illustrated in Sec.\ IVD of CF94, effects due to prior information and
which scale nonlinearly with $1/\mbox{SNR}$ contribute significantly
at low SNR.  The Gaussian approximation then tends to underestimate
measurement errors by missing tails or multimodal structure in
posterior distributions.

We emphasize that in this analysis we do not consider systematic
errors that occur due to limitations in our source model or to
gravitational lensing effects.  A framework for analyzing systematic
errors in GW measurements has recently been presented by \cite{cv07}.
An important follow-on to this work will be to estimate systematic
effects and determine whether they significantly change our
conclusions.

\subsection{Binary Selection and Priors}
\label{sec:selectionandpriors}

We now describe how we generate a sample of detectable GW-SHB events.
We assume a constant comoving density (\citealt{peebles93},
\citealt{hogg99}) of GW-SHB events, in a $\Lambda$CDM Universe with
$H_0=70.5\ \mbox{km}/\mbox{sec}/\mbox{Mpc}$, $\Omega_{\Lambda}=0.726$,
and $\Omega_{m}=0.2732$ \citep{komatsu09}.  We distribute $10^6$
binaries uniformly in volume with random sky positions and
orientations to redshift $z = 1$ ($D_L \simeq 6.6$ Gpc).  We then
compute the average SNR, Eq.\ (\ref{eq:snr_ave}), for each binary at
each detector, and use Eq.\ (\ref{eq:snr_tot_ave}) to compute the
average total SNR for each network we consider.  We assume prior
knowledge of the merger time (since we have assumed that the inspiral is
correlated with a SHB), so we set a threshold SNR for the {\it total}
detector network, $\mbox{SNR}_{\rm total} = 7.5$ (see discussion in
DHHJ06).  This is somewhat reduced from the threshold we would set in
the absence of a counterpart, since prior knowledge of merger time and
source position reduces the number of search templates we need by a
factor $\sim 10^{5}$ (\citealt{kp93}, \citealt{Owen96}).  Using the
average SNR to set our threshold introduces a slight error into our
analysis, since the true SNR will differ from the average. Some events
which we identify as above threshold could be moved below threshold
due to a measurement's particular noise realization.  However, some
sub-threshold events will likewise be moved above threshold, and the net
effect is not expected to be significant.

\begin{figure}
\centering 
\includegraphics[width=0.9\columnwidth]{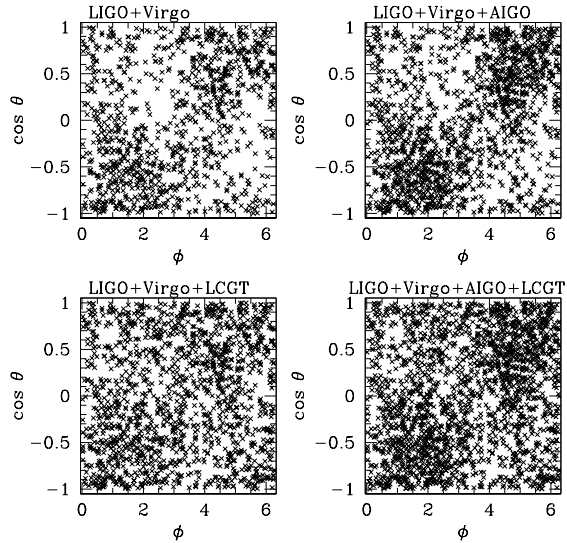}
\caption{Detected NS-NS binaries for our various detector networks as
a function of sky position $(\cos\theta,\phi)$.  The lower right panel
shows the binaries detected by a five-detector network (both LIGO
sites, Virgo, AIGO, and LCGT).  We find that LIGO plus Virgo (our
``base'' network) only detects $50\%$ of the five-detector events;
LIGO, Virgo, and AIGO detect $74\%$ of these events; and LIGO, Virgo,
and LCGT, detect $72\%$ of these events.  Detections are more
uniformly distributed on the sky in networks that include LCGT; AIGO
improves coverage in two of the sky's quadrants.  Our coordinate
$\phi$ is related to right ascension $\alpha$ by $\phi=\alpha-$GMST,
where GMST is Greenwich Mean Sidereal Time; $\theta$ is related to
declination $\delta$ by $\theta = \pi/2 - \delta$.}
\label{fig:detected_binaries}
\end{figure}

Our threshold selects detectable GW-SHB events for each detector
network.  We define ``total detected binaries'' to mean binaries which
are detected by a network of all five detectors---both LIGO
sites, Virgo, AIGO, and LCGT.  Including AIGO and LCGT substantially
improves the number detected, as compared to just using the two LIGO
detectors and Virgo.  Assuming that all binary orientations are
equally likely given an SHB (i.e., no beaming), we find that a LIGO-Virgo
network detects 
$50\%$ of the total detected binaries; LIGO-Virgo-AIGO detects $74\%$
of the total; and LIGO-Virgo-LCGT detects $72\%$ of the total.  Figure
\ref{fig:detected_binaries} shows the sky distribution of detected binaries for
various detector combinations.
Networks which include LCGT tend to have rather uniform sky coverage.
Those with AIGO cover the quadrants $\cos\theta > 0$, $\phi > \pi$ and
$\cos\theta < 0$, $\phi < \pi$ particularly well.

\begin{figure}
\centering 
\includegraphics[width=0.98\columnwidth]{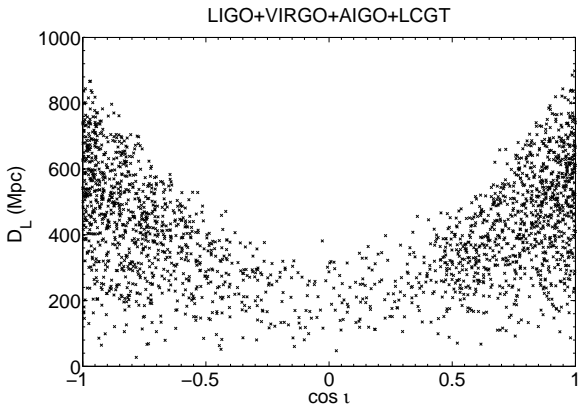}
\caption{The 2-D marginalized prior distribution in luminosity
distance $D_L$ and cosine inclination $\cos \iota$.  Each point
represents a detected NS-NS binary for a network comprising all five
detectors.  Notice the bias toward detecting face-on binaries
($\cos\iota \to \pm 1$)---they are detected to much larger distances
than edge-on ($\cos\iota \to 0$).}
\label{fig:marg2DDLcosinc}
\end{figure}

\begin{figure}
\centering 
\includegraphics[width=0.98\columnwidth]{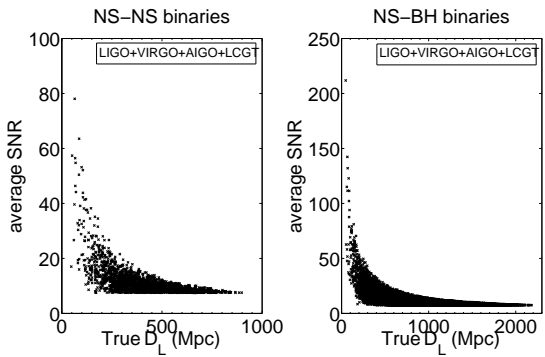}
\caption{Average network SNR versus luminosity distance of the total detected
NS-NS and NS-BH binaries. This assumes an idealized network consisting
of both LIGO detectors,
Virgo, AIGO, and LCGT.  Left panel shows all detected NS-NS binaries
(one point with SNR above 100 is omitted); right panel shows all
detected NS-BH binaries (one point with SNR above 350 is omitted).
Notice the different axis scales: NS-BH binaries are detected to more
than twice the distance of NS-NS. The threshold SNR for the {\it total}
detector network is 7.5, $\mbox{SNR}_{\rm total} = 7.5$.}
\label{fig:SNRvsDL}
\end{figure}

Our selection method implicitly sets a prior distribution on our
parameters.  For example, the thresholding procedure results in a
significant bias in detected events toward face-on binaries, with
$\mathbf{\hat L}\cdot \mathbf{\hat n} \rightarrow \pm 1$.  Figure
{\ref{fig:marg2DDLcosinc}} shows the distribution of detectable NS-NS
binaries for the parameters $\left(\cos\iota, D_L\right)$.  Since we
use an unrealistic mass distribution $(1.4\,M_\odot$--$1.4\,M_\odot$
NS-NS and $1.4\,M_\odot$--$10\,M_\odot$ NS-BH binaries), instead of a
more astrophysically realistic distribution, the implicit mass prior
is uninteresting.  Figure \ref{fig:SNRvsDL} shows the average total
SNR versus the true $D_L$ of our sample of detectable NS-NS and NS-BH
binaries for our ``full'' network (LIGO, Virgo, AIGO, LCGT).  Very few
detected binaries have SNR above 30 for NS-NS, and above 70 for NS-BH.
It is interesting to note the different detectable ranges between the
two populations: NS-BH binaries are detectable to over twice the
distance of NS-NS binaries.

We are also interested in seeing the impact that prior knowledge of
SHB collimation may have on our ability to measure these events.  To
date there exist only two tentative observations which suggest that
SHBs may be collimated (\citealt{grupeetal06},
\citealt{burrowsetal06}, \citealt{soderbergetal06}); we therefore
present results for moderate collimation and for isotropic SHB
emission.  To obtain a sample of beamed SHBs, we assume that the burst
emission is collimated along the orbital angular momentum axis, where
baryon loading is minimized.  Following DHHJ06, we use a distribution
for $\cos\iota \equiv v$ of $dP/dv \propto \exp[-(1 -
v)^2/2\sigma_v^2]$, with $\sigma_v=0.05$.  This corresponds to a
beamed population with $68\%$ of its distribution having an opening
jet angle within roughly $25^\circ$.  We construct a beamed subsample
by selecting events from the total sample of detected events such that
the final distribution in inclination angle follows $dP/dv$.  Joint
measurements of SHBs and GW-driven inspirals should enable us to
constrain beaming angles by comparing the measured rates for these two
populations.

\subsection{Markov-Chain Monte-Carlo approach}
\label{sec:mhmc}

The principle disadvantage of the Bayes estimators
$\tilde\theta^i_{\rm BAYES}$ and $\Sigma^{ij}_{\rm BAYES}$ is the high
computational cost of evaluating the multi-dimensional integrals which
define them, Eqs.\ (\ref{eq:Bayes}) and (\ref{eq:sigma_bayes}).  To
get around this problem, we use Markov-Chain Monte-Carlo (MCMC)
methods to explore the PDFs describing the seven parameters $\{ {\cal
M}_c, \mu, D_L, \cos \iota, \psi, t_c, \Phi_c \}$.  MCMC methods are
widely used in diverse astrophysical applications, ranging from high
precision cosmology (e.g.\ \citealt{dunkley09}, \citealt{sievers09})
to extra-solar planet studies (e.g.\ \citealt{ford05},
\citealt{winn07}).  They have seen increased use in GW measurement and
parameter estimation studies in recent years (e.g.,
\citealt{stroeer06}, \citealt{wickham06}, \citealt{cornish07},
\citealt{porter08}, \citealt{rover07}, \citealt{vandersluys08}).

MCMC generates a random sequence of parameter states that sample the
posterior distribution, $p(\boldsymbol{\theta} | \mathbf{s})$.  Let
the $n$th sample in the sequence be $\boldsymbol{\theta}^{(n)}$.  If
one draws a total of $N$ random samples, Eqs.\ (\ref{eq:Bayes}) and
(\ref{eq:sigma_bayes}) can then be approximated as sample averages:
\begin{eqnarray}
{\tilde\theta}^i_{\rm BAYES} &\simeq& \frac{1}{N} \sum_{n = 1}^N
(\theta^i)^{(n)}\;,
\label{eq:approxBayes}\\
\Sigma^{ij}_{\rm BAYES} &\simeq& \frac{1}{N} \sum_{n = 1}^N
\left(\tilde\theta^i_{\rm BAYES} - (\theta^i)^{(n)}\right)
\left(\tilde\theta^j_{\rm BAYES} - (\theta^j)^{(n)}\right)\;.
\nonumber\\
\label{eq:approxsigma_bayes}
\end{eqnarray}
The key to making this technique work is drawing a sequence that
represents the posterior PDF.  We use the Metropolis-Hastings
algorithm to do this (\citealt{metropolis53}, \citealt{hastings70});
see \cite{neal93}, \cite{gilks96}, \cite{mackay03}, and \cite{cml04}
for in-depth discussion.  The MCMC algorithm we use is based on a
generic version of CosmoMC\footnote{See
http://cosmologist.info/cosmomc/}, described in \cite{lewis02}.

Appropriate priors are crucial to any MCMC analysis.  We take the
prior distributions in chirp mass ${\cal M}_z$, reduced mass $\mu_z$,
polarization angle $\psi$, coalescence time $t_c$, and coalescence
phase $\Phi_c$ to be {\it flat} over the region of sample space where
the binary is detectable according to our selection procedure.  More
specifically, we choose
\begin{itemize}

\item $p^{(0)}({\cal M}_z) = {\rm constant}$ over the range
$[1\,M_\odot$, $2\,M_\odot]$ for NS-NS; and over the range
$[2.5\,M_\odot$, $4.9\,M_\odot]$ for NS-BH.  (The true chirp masses in
the binaries' rest frames are $1.2\,M_\odot$ for NS-NS and
$3.0\,M_\odot$ for NS-BH.)

\item $p^{(0)}(\mu_z) = {\rm constant}$ over the range
$[0.3\,M_\odot$, $2\,M_\odot]$ for NS-NS; and over the range
$[0.5\,M_\odot$, $3.5\,M_\odot]$ for NS-BH.  (The true reduced masses
in the binaries' rest frames are $0.7\,M_\odot$ for NS-NS and
$1.2\,M_\odot$ for NS-BH.)

\item $p^{(0)}(\psi) = {\rm constant}$ over the range $[0,\pi]$.

\item $p^{(0)}(t_c) = {\rm constant}$ over the range $[-100\,{\rm
sec}, 100\,{\rm sec}]$.  Since we assume that $t_c$ is close to the
time of the SHB event, it is essentially the time offset between the
system's final GWs and its SHB photons.  We find that the range in
$t_c$ we choose is almost irrelevant, as long as the prior is flat and
includes the true value.  No matter how broad we choose the prior in
$t_c$, our posterior PDF ends up narrowly peaked around $\hat t_c$.

\item $p^{(0)}(\Phi_c) = {\rm constant}$ over the range $[0, 2\pi]$.

\end{itemize}
The prior distribution for $D_L$ is inferred by taking the density of
SHBs to be uniform per unit comoving volume over the luminosity
distance range [0, 2 Gpc] for NS-NS binaries, and over the range [0, 5
Gpc] for NS-BH binaries.  For our sample with isotropic inclination
distribution, we put $p^{(0)}(\cos \iota) = {\rm constant}$ over the
range $[-1,1]$.  When we assume SHB collimation, our prior in
$\cos\iota \equiv v$ is the same as the one that we used in our
selection procedure discussed previously:
\begin{equation}
\frac{dp^{(0)}}{dv}(v) \propto e^{-(1 -v)^2/2\sigma_v^2}\;,
\end{equation}
with $\sigma_v = 0.05$.

We then map out full distributions for each of our seven parameters,
assessing the mean values [Eq.\ (\ref{eq:Bayes})] and the standard
deviations [Eq.\ (\ref{eq:sigma_bayes})].  We generate four chains
which run in parallel on the CITA ``Sunnyvale'' Cluster.  Each chain
runs for a maximum of $10^7$ steps; we find that the mean and median
number of steps are $\sim 10^5$ and $\sim 10^4$, respectively. Each
evaluation of the likelihood function takes $\sim0.3$ seconds.  We use
the first 30\% of a chain's sample states for ``burn in,'' and
discard that data.  Our chains start at random offset parameter
values, drawn from Gaussians centered on the true parameter values.  We
assess convergence by testing whether the multiple chains have
produced consistent parameter distributions.  Following standard
practice, we use the Gelman-Rubin convergence criterion, defining a
sequence as ``converged'' if the statistic $R < 1.1$ on the last half
of our samples; see \cite{gr92} for more details. We use convergence
as our stopping criterion.  Each simulation for every binary runs for
an hour to forty-eight hours; the mean and median runtime are eight
and three hours, respectively.

\subsection{The ``averaged'' posterior PDF}
\label{sec:averagedPDF}

Central to the procedure outlined above is the use of the datastream
${\bf s} = {\bf h}(\boldsymbol{\theta}) + {\bf n}$ which enters the
likelihood function ${\cal L}_{\rm TOT}({\bf s} |
\boldsymbol{\theta})$.  The resulting posterior PDF, and the
parameters one infers, thus depend on the noise ${\bf n}$ which one
uses.  One may want to evaluate statistics that are in a well-defined
sense ``typical'' given the average noise properties, rather than
depending on a particular noise instance.  Such averaging is
appropriate, for example, when forecasting how well an instrument
should be able to measure the properties of a source or process.  We
have also found it is necessary to average when trying to compare our
MCMC code's output with previous work.

As derived below, the averaged posterior PDF takes a remarkably simple
form: It is the ``usual'' posterior PDF, Eq.\ (\ref{eq:postPDF}) with
the noise ${\bf n}$ set to {\it zero}.  This does not mean that one
ignores noise when constructing the averaged PDF; one still relates
signal amplitude to typical noise by the average SNR, Eq.\
(\ref{eq:snr_ave}).  As such, the averaged statistics will show an
improvement in measurement accuracy as SNR is increased.

To develop a useful notion of averaged posterior PDF, consider the
hypothetical (and wholly unrealistic) case in which we measure a
signal using $M$ different noise realizations for the same event.  The
joint likelihood for these measurements is
\begin{equation}
{\cal L}^{\rm joint}_{\rm TOT}({\bf s}_1, {\bf s}_2, \ldots {\bf
s}_M | \boldsymbol{\theta} ) = \prod_{i=1}^M {\cal L}_{\rm TOT} ({\bf s}_i | \boldsymbol{\theta})\;.
\label{eq:pjointdef}
\end{equation}
Let us define the ``average'' PDF as the product of the prior
distribution of the parameters multiplied by the geometric mean of the
likelihoods which describe these measurements:
\begin{equation}
p_{\rm ave}(\boldsymbol{\theta} | {\bf s}) \equiv  {\cal N} \, p^{(0)} 
{\cal L}^{\rm joint}_{\rm TOT}({\bf s}_1, {\bf s}_2, \ldots {\bf
s}_M| \boldsymbol{\theta})^{1/M}\;.
\label{eq:pavedef}
\end{equation}
Expanding this definition, we find
\begin{eqnarray}
p_{\rm ave} (\boldsymbol{\theta} | \bf{s}) & \equiv &   {\cal N} \,
p^{(0)} \prod_{i=1}^{M}\, \left[  {\cal L}_{\rm TOT} ({\bf s}_i |
  \boldsymbol{\theta}) \right]^{1/M}\;, 
\label{eq:avepostPDF}
\end{eqnarray}
where the subscript $i$ denotes the $i$th noise realization in our set
of $M$ observations.  The ``ensemble average likelihood function'' can
in turn be expanded as
\begin{eqnarray}
\prod_{i=1}^{M} \left[{\cal L}_{\rm TOT} ({\bf s}_i |
{\boldsymbol{\theta}})\right]^{1/M} & = & \prod_{a} \prod_{i=1}^{M}
\left[{\cal L}_{a} (s_{a,i} \, | \,{\boldsymbol \theta})\right]^{1/M}
\nonumber\\
&=& \prod_{a} \prod_{i=1}^{M} e^{ - \big(h_a({\boldsymbol
 \theta}) - s_{a,i} \, \big| \, h_a({\boldsymbol \theta}) - s_{a,i}
 \big)/2M }
\nonumber\\
& = & \prod_{a} e^{-\big( h_a({\boldsymbol \theta}) -
 h_a(\boldsymbol{\hat{\theta}}) \, \big| \, h_a({\boldsymbol \theta})
 - h_a (\boldsymbol{\hat{\theta}})\big)/2}
\nonumber\\
&\times&
\prod_{i=1}^M \exp\left[\frac{1}{M}\left( n_{a,i} \, \bigg|
 h_{a}(\boldsymbol{\theta}) - h_{a}(\boldsymbol{\hat\theta})\right)
 \right]
\nonumber\\
&\times&
\prod_{i=1}^M \exp\left[-\frac{1}{2M}\left( n_{a,i} \, \bigg|
n_{a,i}\right)\right]\;.
\label{eq:multi_obs_likelihood}
\end{eqnarray}
By taking $M$ to be large, the last two lines of Eq.\
(\ref{eq:multi_obs_likelihood}) can be evaluated as follows:
\begin{eqnarray}
& &\prod_{i=1}^M \exp\left[\frac{1}{M}\left( n_{a,i} \, \bigg|
 h_{a}(\boldsymbol{\theta}) - h_{a}(\boldsymbol{\hat\theta})\right)
 \right] 
\nonumber\\
& &\qquad\qquad = \exp\left[\frac{1}{M}\sum_{i = 1}^M \left(
  n_{a,i} \, \bigg| h_{a}(\boldsymbol{\theta}) -
  h_{a}(\boldsymbol{\hat\theta})\right) \right]
\nonumber\\
& &\qquad\qquad \simeq \exp\left[\left\langle \left( n_{a} \,
 \bigg| h_{a}(\boldsymbol{\theta}) -
 h_{a}(\boldsymbol{\hat\theta})\right)\right\rangle \right]
\nonumber\\
& & \qquad\qquad = 1\;.
\end{eqnarray}
Here, $\langle \ldots \rangle$ denotes an ensemble average over noise
realizations (cf.\ Sec.\ {\ref{sec:formaloverview}}), and we have used
the fact that our noise has zero mean.  Similarly, we find
\begin{eqnarray}
\prod_{i=1}^M \exp\left[-\frac{1}{2M}\left( n_{a,i} \, \bigg|
 n_{a,i}\right)\right] &=& \exp\left[-\frac{1}{2M}\sum_{i = 1}^M
 \left( n_{a,i} \, \bigg| n_{a,i}\right) \right]
\nonumber\\ &\simeq&
 \exp\left[-\frac{1}{2}\left\langle \left( n_{a} \, \bigg|
 n_{a}\right)\right\rangle \right]
 \nonumber\\ &=& e^{-1}\;.
\end{eqnarray}
This uses $\langle (n_a | n_a) \rangle = 2$, which can be proved using
the noise properties (\ref{eq:autocov}), (\ref{eq:sn_def}), and
(\ref{eq:noisestats}).

Putting all this together, we finally find
\begin{equation}
p_{\rm ave}(\boldsymbol{\theta} | {\bf s}) = {\cal N}
 p^{0}(\boldsymbol{\theta}) \prod_{a} e^{ -\big( h_a({\boldsymbol
 \theta}) - h_a (\boldsymbol{\hat{\theta}}) \, \big| \,
 h_a({\boldsymbol \theta}) - h_a (\boldsymbol{\hat{\theta}}) \big)/2
 }\;,
\label{eq:avepostPDF_final2}
\end{equation}
where we have absorbed $e^{-1}$ into the normalization ${\cal N}$.
The posterior PDF, averaged over noise realizations, is simply
obtained by evaluating Eq.\ (\ref{eq:postPDF}) with the noise ${\bf
n}$ set to zero.

\section{Results I: Validation and Testing}
\label{sec:valid}

We now validate and test our MCMC code against results from CF94.  In
particular, we examine the posterior PDF for the NS-NS binary which was
studied in detail in CF94.  We also explore the dependence of distance
measurement accuracies on the detector network and luminosity
distance, focusing on the strong degeneracy that exists between $\cos
\iota$ and $D_L$.

\subsection{Comparison with CF94}
\label{sec:cf}

Validation of our MCMC results requires comparing to work which goes
beyond the Gaussian approximation and Fisher matrix estimators.  In
Section IVD of CF94, Cutler \& Flanagan investigate effects that are
non-linear in $1/\mbox{SNR}$.  They show that such effects
have a significant impact on distance measurement accuracies for low
SNR.  In particular, they find that Fisher-based estimates understate
distance measurement errors for a network of two LIGO detectors
and Virgo.

Because they go beyond a Fisher matrix analysis, the results of CF94
are useful for comparing to our results.  Their paper is also useful
in that they take source position to be known.  Our approach is
sufficiently different from CF94 that we do not expect perfect
agreement, however.  The most important difference is that we directly map out
the posterior PDF and compute sample averages using Eqs.\
(\ref{eq:Bayes}) and (\ref{eq:sigma_bayes}), for the full parameter
set $\{ {\cal M}_z, \mu_z, D_L, \cos \iota, \psi, t_c, \Phi_c \}$.  In
contrast, CF94 estimate measurement errors only for $D_L$, using an
approximation on an analytic Bayesian derivation of the marginalized
PDF for $D_L$.  Specifically, Cutler \& Flanagan expand the
exponential factor in Eq.\ (\ref{eq:postPDF}) beyond second order in
terms of some ``best-fit'' maximum likelihood parameters.  Their
approximation treats strong correlations between the parameters $D_L$
and $\cos \iota$ that are non-linear in 1/SNR.  However, other
correlations between $D_L$ and $(\psi, \phi_c)$ are only considered to
linear order. They obtain an analytic expression for the posterior PDF
of the variables $D_L$ and $\cos \iota$ in terms of their ``best-fit''
maximum-likelihood values $\tilde{D}_L$ and $\cos \tilde{\iota}$ [see
Eq.\ (4.57) of CF94]. The marginalized 1-D posterior PDFs for $D_L$
are then computed by numerically integrating over $\cos \iota$.  The
1-D marginalized PDF we compute in parameter $\theta_i$ is
\begin{equation}
\label{eq:margPDF}
p_{\rm marg}(\theta_i | {\bf s}) = \int \dots \int
p(\boldsymbol{\theta} | {\bf s}) d\theta_1 \dots d\theta_{i-1}\;
d\theta_{i+1} \dots d\theta_N
\end{equation}
where $p(\boldsymbol{\theta} | \bf{s})$ is the posterior PDF given by
Eq.\ (\ref{eq:postPDF}) and $N$ is the number of dimensions of our
parameter set.

In addition to this rather significant difference in techniques, there
are some minor differences which also affect our comparison:

\begin{itemize}

\item We use the restricted 2PN waveform; CF94 use the leading
``Newtonian, quadrupole'' waveform that we used for pedagogical
purposes in Sec.\ \ref{sec:sirens}.  Since distance is encoded in the
waveform's amplitude, we do not expect that our use of a higher-order
phase function will have a large impact.  However, to avoid any easily
circumvented mismatch, we adopt the Newtonian-quadrupole waveform for
these comparisons.  This waveform does not depend on reduced mass
$\mu$, so {\it for the purpose of this comparison only}, our parameter
space is reduced from 7 to 6 dimensions.

\item We use the projected advanced sensitivity noise curve shown in
Fig.\ (\ref{fig:aligonoise}); CF94 use an analytical form [their Eq.\
(2.1)\footnote{Note that it is missing an overall factor of $1/5$ (E.\
E.\ Flanagan, private communication).}] based on the best-guess for
what advanced sensitivity would achieve at the time of their analysis.
Compared to the most recent projected sensitivity, their curve
underestimates the noise at middle frequencies ($\sim 40$--$150$ Hz)
and overestimates it at high frequencies ($\gtrsim 200$ Hz).  We adopt
their noise curve for this comparison.  Because of these differences,
CF94 rather seriously overestimates the SNR for NS-NS inspiral.  Using
their noise curve, the average SNR for the binary analyzed in their Fig.\
10 is 12.4\footnote{CF94 actually report an SNR of 12.8.  The
discrepancy is due to rounding the parameter $r_0$ in their Eq.\
(4.28).  Adjusting to their preferred value (rather than computing
$r_0$) gives perfect agreement.}; using our up-to-date model for
advanced LIGO, it is 5.8.  As such, the reader should view the numbers
in this section of our analysis as useful {\it only} for validation
purposes.

\item The two analyses use different priors. As extensively discussed
in Sec.\ {\ref{sec:mhmc}}, we set uniform priors on the chirp mass
${\cal M}_z$, on the time $t_c$ and phase $\Phi_c$ at coalescence, and
on the polarization phase $\psi$.  For this comparison, we assume
isotropic emission and set a flat prior on $\cos\iota$.  We
assume our sources are uniformly distributed in constant comoving
volume.  However, our detection threshold depends on the total network
SNR, and effectively sets a joint prior on source inclination and
distance. CF94 use a prior distribution only for the set $\{ D_L, \cos
\iota, \psi, \Phi_c \}$ that is flat in polarization phase,
coalescence phase, and inclination.  They assume a prior that is
uniform in volume, but that cuts off the distribution at a distance
$D_{L,{\rm max}} \simeq 6.5\,{\rm Gpc}$.

\end{itemize}

Our goal here is to reproduce the 1-D marginalized posterior PDF in
$D_L$ for the binary shown in Fig.\ 10 of CF94.  We call this system
the ``CF binary.''  Each NS in the CF binary has $m_z = 1.4\,M_\odot$ and
sky position $(\theta, \phi) = (50^{\circ}, 276^{\circ})$;
the detector network comprises LIGO Hanford, LIGO Livingston and
Virgo.  CF94 report the ``best-fit'' maximum-likelihood values
($\tilde{D}_L$, $\cos\tilde{\iota}$, $\tilde{\Psi}$) to be ($432\,{\rm
Mpc}$, $0.31$, $101.5^{\circ}$), where $\Psi = \psi + \Delta \psi
({\bf n})$, and where $\Delta \psi ({\bf n})$ depends on the preferred
basis of ${\bf e}^{\times}$ and ${\bf e}^{\times}$ set by the detector
network [see Eqs.\ (4.23)--(4.25) of CF94\footnote{Note that Eq.\
(4.25) of CF94 should read $\tan(4\Delta\psi) =
2\Theta_{+\times}/(\Theta_{++} - \Theta_{\times\times})$. In addition,
$\tilde\Psi = 56.5^{\circ}$ should read $\tilde\Psi = 101.5^{\circ}$
under the caption of Fig.\ 10.  (We have changed notation from
$\bar\psi$ in CF94 to $\Psi$ to avoid multiple accents on the best fit
value.)  We thank \'Eanna Flanagan for confirming these
corrections.}].  To compare our distribution with theirs, we assume
that $\boldsymbol{\hat\theta} = \boldsymbol{\tilde\theta}_{\rm ML}$
for the purpose of computing the likelihood function ${\cal
L}(\boldsymbol{\theta} | {\bf s})$.  This is a reasonable assumption
when the priors are uniform over the relevant parameter space.  As
already mentioned, for this comparison we use their advanced detector
noise curve and the Newtonian-quadrupole waveform.  Finally, we
interpret the solid curve in Fig.\ 10 of CF94 as the marginalized 1-D
posterior PDF in $D_L$ for an average of posterior PDFs of parameters
(given an ensemble of many noisy observations for a particular event).
We compute the average PDF as described in Sec.\
{\ref{sec:averagedPDF}}, and then marginalize over all parameters
except $D_L$, using Eq.\ (\ref{eq:margPDF}).

\begin{figure}
\centering 
\includegraphics[width=1.00\columnwidth]{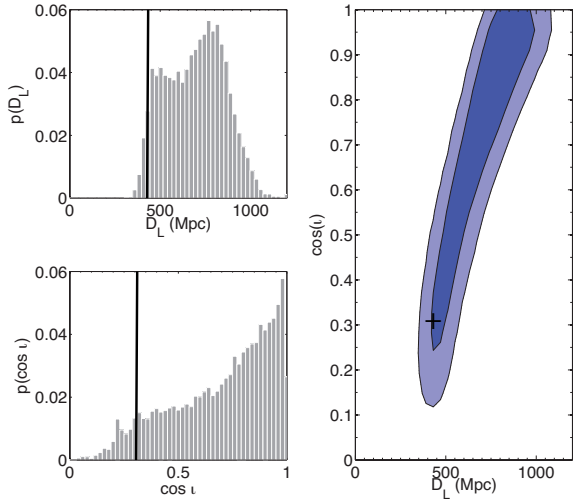}
\caption{1-D and 2-D marginalized posterior PDFs for $D_L$ and
$\cos \iota$ averaged over noise (as described in Sec.\
\ref{sec:averagedPDF}) for the ``CF binary.''  Our goal is to
reproduce, as closely as possible, the non-Gaussian limit summarized
in Fig.\ 10 of CF94.  Top left panel shows the 1-D marginalized
posterior PDF in $D_L$ (the true value $\hat D_L = 432\,{\rm Mpc}$ is
marked with a solid black line); bottom left panel illustrates the 1-D
marginalized posterior PDF in $\cos \iota$ (true value $\cos \hat
\iota = 0.31$ likewise marked).  The right-hand panel shows the 2-D
marginalized posterior PDF for $D_L$ and $\cos \iota$; the true values
($\hat D_L = 432\,{\rm Mpc}, \cos \hat \iota = 0.31$) are marked with a cross.  The contours
around the dark and light areas indicate the 68 and 95\% interval
levels, respectively.  The true values lie within the 68\% interval.
The Bayes mean and rms measurement accuracies are (694.4 Mpc, 0.70)
and (162 Mpc, 0.229) for ($D_L$, $\cos \iota$), respectively.}
\label{fig:CFbinary}
\end{figure}

The left-hand panels of Fig.\ \ref{fig:CFbinary} show the resulting
1-D marginalized PDF in $D_L$ and $\cos \iota$.  Its shape has a broad
structure not dissimilar to the solid curve shown in Fig.\ 10 of CF94:
The distribution has a small bump near $D_L \approx 460\,{\rm Mpc}$, a
main peak at $D_L \approx 700\,{\rm Mpc}$, and extends out to roughly
1 Gigaparsec.  Because of the broad shape, the Bayes mean ($\tilde
D_{L,\rm{BAYES}} = 694\,{\rm Mpc}$) is significantly different from
both the true value ($\hat D_L = 432\,{\rm Mpc}$ in our calculation)
and from the maximum likelihood ($\tilde D_{L,\rm{ML}} = 495\,{\rm
Mpc}$).  Thanks to the marginalization, the peak of this curve does
not coincide with the maximum likelihood.
    
We further determine the 2-D marginalized posterior PDFs in $D_L$ and
$\cos \iota$ for the CF binary.  Figure \ref{fig:CFbinary} illustrates
directly the very strong degeneracy between these parameters, as
expected from the form of Eqs.\ (\ref{eq:hplus}) and
(\ref{eq:hcross}), as well as from earlier works (e.g.,
\citealt{markovic93}, CF94). It's worth noting that, as CF94 comment,
this binary is measured particularly poorly.  This is largely due to
the fact that one polarization is measured far better than the other,
so that the $D_L$--$\cos\iota$ degeneracy is essentially unbroken.
This degeneracy is responsible for the characteristic tail to large
$D_L$ we find in the 1-D marginalized posterior PDF in $D_L$, $p(D_L |
\bf{s})$, which we investigate further in the following section.

\subsection{Test 1: Varying luminosity distance and number of detectors}
\label{sec:cf_vary}

\begin{figure}
\centering 
\includegraphics[width=1.1\columnwidth]{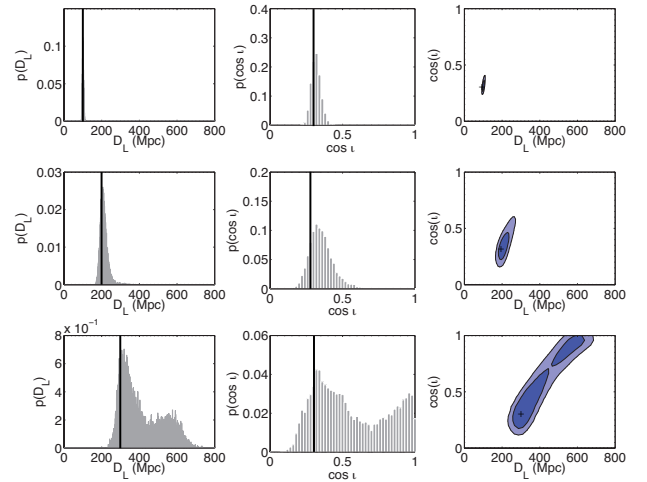}
\caption{1-D and 2-D marginalized PDFs for $D_L$ and $\cos \iota$,
averaged (as described in Sec.\ \ref{sec:averagedPDF}) over noise
ensembles for the ``CF binary'' at different values of true luminosity
distance $\hat{D}_L$: [100 Mpc, 200 Mpc, 300 Mpc] (top to bottom).
True parameter values are marked with a solid black line or a black
cross.  The Bayes means and rms errors on luminosity distance are
[101.0 Mpc, 212.1 Mpc, 411.2 Mpc] and [3.6 Mpc, 21.4 Mpc, 110.0 Mpc],
respectively.  The corresponding means and errors for $\cos \iota$ are
[0.317, 0.357, 0.562] and [0.033, 0.089, 0.247].  The dark and light
contours in the 2-D marginalized PDF plots indicate the 68 and 95\%
interval levels, respectively.  The true value always lies within the
68\% contour region of the 2-D marginalized area at these distances.}
\label{fig:CFbinaryvaryingDLa}
\end{figure}

\begin{figure}
\centering 
\includegraphics[width=1.1\columnwidth]{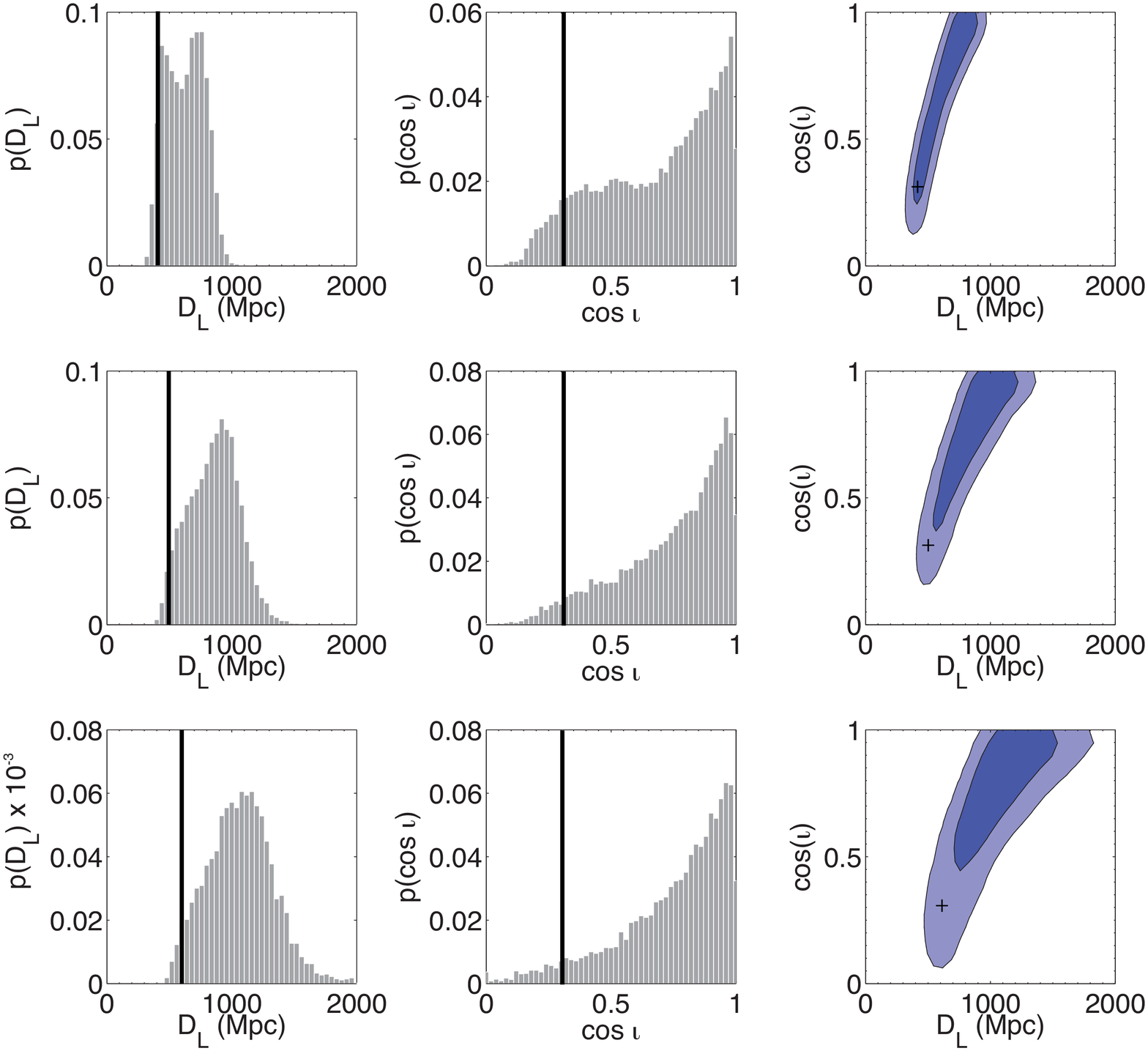}
\caption{Same as Fig.\ {\ref{fig:CFbinaryvaryingDLa}}, but for true
luminosity distance $\hat{D}_L =$ [400 Mpc, 500 Mpc, 600 Mpc] (top to
bottom).  True parameter values are marked with a solid black line or a black
cross. In this case, the Bayes means and rms errors for luminosity
distance are [627.17 Mpc, 857.3 Mpc, 1068 Mpc] and [148.8 Mpc, 198.1
Mpc, 262.2 Mpc], respectively.  The means and errors for $\cos \iota$
are [0.686, 0.745, 0.746] and [0.237, 0.209, 0.218].  The dark and
light contours in the 2-D marginalized PDF plots indicate the 68 and
95\% interval levels, respectively.  The true value lies within the
68\% contour region for $D_L = 400$ Mpc, but moves outside this region
for larger values.}
\label{fig:CFbinaryvaryingDLb}
\end{figure}

We now examine how well we measure $D_L$ as a function of distance to
the CF binary and the properties of the GW detector network.  Figures
\ref{fig:CFbinaryvaryingDLa} and \ref{fig:CFbinaryvaryingDLb} show the
1-D and 2-D marginalized posterior PDFs in $D_L$ and $\cos \iota$ for
the CF binary at $\hat{D}_L = \{100$, $200$, $300$, $400$, $500$,
$600\}$ Mpc.  For all these cases, we keep the binary's sky position,
inclination, and polarization angle fixed as in Sec.\ \ref{sec:cf}.
The average network SNRs we find for these six cases are (going from
$\hat D_L = 100\,{\rm Mpc}$ to $600\,{\rm Mpc}$) 53.6, 26.8, 17.9,
13.4, 10.7, and 8.9 (scaling as $1/\hat D_L$).  Interestingly, the
marginalized PDFs for both distance and $\cos\iota$ shown in Figs.\
{\ref{fig:CFbinaryvaryingDLa}} and {\ref{fig:CFbinaryvaryingDLb}} have
fairly Gaussian shapes for $\hat D_L = 100$ and 200 Mpc, but have very
non-Gaussian shapes for $\hat D_L \ge 300\,{\rm Mpc}$.  This can be
considered ``anecdotal'' evidence that the Gaussian approximation for
the posterior PDF breaks down at ${\rm SNR} \lesssim 25$ or so, at
least for this case.  For lower SNR, the degeneracy between
$\cos\iota$ and $D_L$ becomes so severe that the 1-D errors on these
parameters become quite large.

Next, we examine measurement accuracy versus detector network.  For
the CF binary, adding detectors does not substantially increase the
total SNR.  We increase the average total SNR from 12.4 to 14.6
(adding only AIGO), to 12.4 (adding only LCGT; its contribution is so
small that the change is insignificant to the stated precision), or to
14.7 (adding both AIGO and LCGT).  The average SNR in our detectors is
8.23 for LIGO-Hanford, 8.84 for LIGO-Livingston, 2.91 for Virgo, 8.71
for AIGO, and 1.1 for LCGT.  This pathology is an example of a fairly
general trend that we see; it is common for the SNR to be quite low in
one or more detectors.

In the case of the CF binary, we find that adding detectors does not improve the
measurement enough to break the $D_L$--$\cos\iota$ degeneracy.  The
marginalized PDFs as functions of $D_L$ and $\cos\iota$ remain very
similar to Fig.\ {\ref{fig:CFbinary}}, so we do not show them.  As a
consequence, even with additional detectors, the distance errors
remain large and biased.  The bias is because we tend to find
$\cos\iota$ to be larger than the true (relatively edge-on) value (cf.\ lower left-hand
panel of Fig.\ {\ref{fig:CFbinary}}).  Thanks to the
$D_L$--$\cos\iota$ degeneracy, we likewise overestimate distance.

\subsection{Test 2: Varying source inclination}
\label{sec:faceon_cf}

One of the primary results from the CF binary analysis is a strong
degeneracy between $\cos \iota$ and $D_L$.  As Fig.\
\ref{fig:CFbinary} shows, this results in a tail to large distance in
the 1-D marginalized posterior PDF $p(D_L | \bf{s})$, with a Bayes
mean $\tilde D_L = 694\,{\rm Mpc}$ (compared to $\hat D_L = 432\,{\rm
Mpc}$).  Such a bias is of great concern for using binary sources as
standard sirens.

The CF binary has $\cos\hat\iota = 0.31$, meaning that it is nearly
edge-on to the line of sight.  Hypothesizing that the large tails may
be due to its nearly edge-on nature, we consider a complementary
binary that is nearly face on: We fix all of the parameters to those
used for the CF binary, except for the inclination, which we take to
be $\cos\hat\iota = 0.98$.  We call this test case the ``face-on'' CF
binary.  Changing to a more nearly face-on situation substantially
augments the measured SNR; the average SNR for the face-on CF binary
measured by the LIGO/Virgo base network is 24.3 (versus 12.4 for the
CF binary).  We thus expect some improvement simply owing to the
stronger signal.

\begin{figure}
\centering 
\includegraphics[width=1.1\columnwidth]{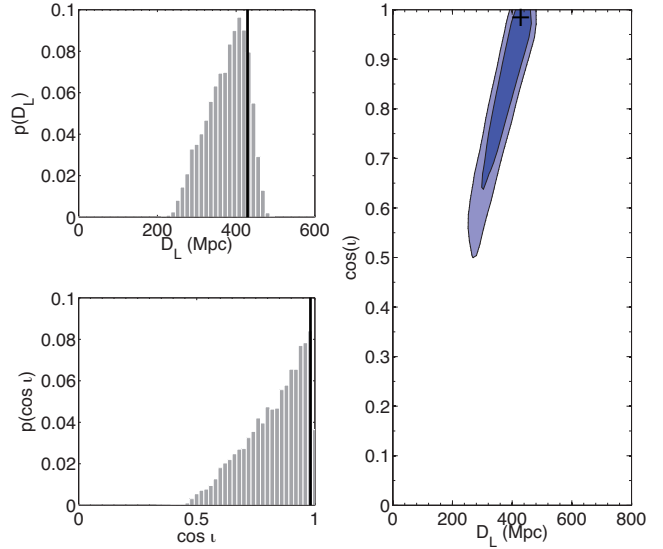}
\caption{Same as Fig.\ {\ref{fig:CFbinary}}, but for the ``face-on''
CF binary.  The Bayes mean and rms errors are (376.3 Mpc, 0.83) and
(51.3 Mpc, 0.12) for ($D_L$, $\cos \iota$), respectively.  Top left
shows the 1-D marginalized posterior PDF in $D_L$ ($\hat D_L =
432\,{\rm Mpc}$ is marked with a solid black line); bottom left shows
the marginalized PDF in $\cos \iota$ (solid black line marks $\cos
\hat \iota = 0.98$).  The right panel shows the 2-D marginalized
posterior PDF; the cross marks the true source parameters ($\hat D_L =
432\,{\rm Mpc}$, $\cos \hat\iota = 0.98$).  As
with the CF binary, the true values lie within the 68\% region.}
\label{fig:SDbinary}
\end{figure}

Figure \ref{fig:SDbinary} shows the 1-D and 2-D marginalized posterior
PDFs in $D_L$ and $\cos \iota$.  As expected, these distributions are
complementary to those we found for the CF binary.  In particular, the
peak of the 1-D marginalized posterior PDF in $D_L$ is shifted to
lower values in $D_L$, and the Bayes mean is much closer to the true
value: $\tilde D_L = 376.3\,{\rm Mpc}$.  The shape of the 1-D
marginalized posterior PDF in $\cos \iota$ is abruptly cut off by the
upper bound of the physical prior $\cos\iota \le 1$, and the tail extends to
lower distances (the opposite of the CF binary).  The Bayes mean
for the inclination is $\cos\tilde\iota = 0.83$.

Just as we varied distance and detector network for the CF binary, we
also do so for the face-on CF binary, with very similar results.  In
particular, varying network has little impact on the marginalized 1-D
PDFs in $D_L$ and $\cos\iota$.  Varying distance, we find that the
marginalized 1-D PDFs are nearly Gaussian in shape for small
distances, but become significantly skewed (similar to the left-hand
panels of Fig.\ {\ref{fig:SDbinary}}) when $\hat D_L > 200$ Mpc.  The
distributions in $\cos\iota$ are particularly skewed thanks to the
hard cut-off at $\cos\iota = 1$.  Interestingly, in this case we tend
to infer a value of $\cos\iota$ that is smaller than the true value.
We likewise find a Bayes mean $\tilde D_L$ that is smaller than $\hat
D_L$.

\subsection{Summary of validation tests}
\label{sec:testing_discuss}

The main result from our testing is that the posterior PDFs we find
have rather long tails, with strong correlations between $\cos\iota$
and $D_L$.  Except for cases with very high SNR, the 1-D marginalized
posterior PDF in $\cos\iota$ tends to be rather broad.  The Bayes mean
for $\cos\iota$ thus typically suggests that a binary is at
intermediate inclination.  As such, we tend to {\it underestimate}
$\cos\iota$ for nearly face-on binaries, and to {\it overestimate} it
for nearly edge-on binaries.  Overcoming this limitation requires us
to either break the $D_L$--$\cos\iota$ degeneracy (such as by setting
a prior on binary inclination), or by measuring a population of
coalescences.  Measuring a population will make it possible to sample
a wide range of the $\cos\iota$ distribution, so that the
event-by-event bias is averaged away in the sample.

\section{Results II: Survey of standard sirens}
\label{sec:main_results}

We now examine how well various detector networks can measure an
ensemble of canonical GW-SHB events.  We randomly choose events from
our sample of {\it detected}\/ NS-NS and NS-BH binaries (where the
selection is detailed in Sec.\ \ref{sec:selectionandpriors}).  We set
a total detector network threshold of 7.5.  Crudely speaking, one
might imagine that this implies, on average, a threshold per detector
of $7.5/\sqrt{5}=3.4$ for a five detector network.  Such a crude ``per
detector threshold'' is useful for getting a rough idea of the range
to which our network can measure events.  Averaging Eq.\
(\ref{eq:snr_ave}) over all sky positions and orientations yields
(DHHJ06)
\begin{eqnarray}
\label{eq:snr_ave_sky_orien}
\left({S\over N}\right)_{a,\ {\rm sky-ave}} &=& \frac{8}{5}
\sqrt{\frac{5}{96}} \frac{c}{D_L} \frac{1}{\pi^{2/3}} \left(\frac{G
{\cal M}_z}{c^3}\right)^{5/6}\times
\nonumber\\
& &\qquad \int_{f_{\rm low}}^{f_{\rm ISCO}}
\frac{f^{-7/3}}{S_h(f)} df\;,
\end{eqnarray}
For total detector network threshold of 7.5, a five detector network
has an average range of about $600\,{\rm Mpc}$ for NS-NS events, and
about $1200\,{\rm Mpc}$ for NS-BH events.  If SHBs are associated with
face-on binary inspiral, these numbers are increased by a factor
$\sqrt{5/2} \simeq 1.58$.  (This factor is incorrectly stated to be
$\sqrt{5/4} \simeq 1.12$ in DHHJ06.)

Let us assume a constant comoving rate of 10 SHBs Gpc$^{3}$ yr$^{-1}$
\citep{nakar06}.  If these events are all NS-NS binary mergers, and
they are isotropically oriented, we expect the full
LIGO-Virgo-AIGO-LCGT network to measure 6 GW-SHB events per year.  If
these events are instead all NS-BH binaries, the full network is
expected to measure 44 events per year.  If these events are beamed,
the factor $1.58$ increases the expected rate to 9 NS-NS or 70 NS-BH
GW-SHB events per year.  We stress that these numbers should be taken
as rough indicators of what the network may be able to measure.
Not all SHBs will be associated with binary inspiral.  Those events
which are will likely include both NS-NS and NS-BH events, with
parameters differing from our canonical choices.  We also do not
account for the fraction of SHBs which will be missed due to
incomplete sky coverage.

In all cases we build our results by constructing the posterior
distribution for an event given a unique noise realization at each
detector.  We keep the noise realization in a given detector and for a
specific binary constant as we add other detectors.  This allows us
to make meaningful comparisons between the performance of different
detector networks.

\subsection{NS-NS binaries}

\begin{figure}
\centering 
\includegraphics[width=1.0\columnwidth]{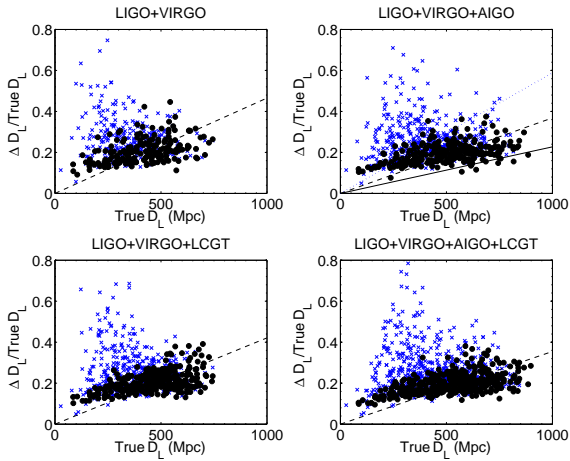}
\caption{Distance measurement errors versus true luminosity distance
for our sample of NS-NS binaries.  Colored crosses assume isotropic
emission; black points assume our beaming prior.  The dashed lines
show the linear best-fit to the beamed sample (see text for
expressions).  In the LIGO+Virgo+AIGO panel we also show the
Fisher-matrix-derived linear scaling given in DHHJ06: $\Delta
D_L/\hat{D}_L \simeq \hat{D}_L/(4.4\,\rm{ Gpc})$ assuming beaming (solid), and
$\Delta D_L/\hat{D}_L \simeq \hat{D}_L/(1.7\,\rm{Gpc})$ for isotropic
emission (dotted).}
\label{fig:DeltaDLNSNS}
\end{figure}

We begin by imagining a population of six hundred detected NS-NS
binaries, either isotropically distributed in inclination angle or
from our beamed subsample, using a network with all five
detectors. Figure \ref{fig:DeltaDLNSNS} shows scatter plots of the
distance measurement accuracies for our unbeamed (blue crosses) and
beamed events (black dots), with each panel corresponding to a
different detector network.  The distance measurement error is defined
as the ratio of the rms measurement error with the true
value\footnote{Our definition differs from that given in CF94, their
Eq.\ (4.62).  Their distance measurement error is described as the
ratio of the rms measurement error with the Bayes mean.  We prefer to
use Eq.\ (\ref{eq:measaccdefn}) as we are interested primarily in the
measurement error given a binary at its true luminosity distance.}
$\hat{D}_L$:
\begin{equation}
\frac{\Delta D_L}{\hat{D}_L} = \frac{\sqrt{\Sigma^{D_L
D_L}}}{\hat{D}_{L}}\;.
\label{eq:measaccdefn}
\end{equation}
$\Sigma^{D_L D_L}$ is computed using (\ref{eq:approxsigma_bayes}).  We
emphasize some general trends in Fig.\ \ref{fig:DeltaDLNSNS} which
are particularly relevant to standard sirens:

\begin{itemize}

\item {\it The unbeamed total sample and the beamed subsample separate
into two distinct distributions.}  As anticipated, the beamed
subsample improves measurement errors in $D_L$ significantly, by
greater than a factor of two or more.  This is due to the beaming
prior, which constrains the
inclination angle, $\cos \iota$, to $\sim 3\%$, thereby breaking the
strong $D_L$--$\cos \iota$ degeneracy.  By contrast, when no beaming
prior is assumed, we find absolute errors of $0.1$--$0.3$ in
$\cos\iota$ for the majority of events.  The strong $D_L$--$\cos
\iota$ degeneracy then increases the distance errors.  A significant
fraction of binaries randomly selected from our sample have $0.5
\lesssim |\cos\hat\iota| < 1$.  As discussed in Sec.\
{\ref{sec:selectionandpriors}}, this is due to the SNR selection
criterion: At fixed distance, face-on binaries are louder and tend to
be preferred.

\item {\it Beamed subsample scalings.}  We fit linear scalings
to our beamed subsample:\\
$\Delta D_L/\hat{D}_L \simeq \hat{D}_L/(2.15\,\rm{Gpc})$ for LIGO + Virgo\\
$\Delta D_L/\hat{D}_L \simeq \hat{D}_L/(2.71\,\rm{Gpc})$ for LIGO + Virgo + AIGO\\
$\Delta D_L/\hat{D}_L \simeq \hat{D}_L/(2.38\,\rm{Gpc})$ for LIGO + Virgo + LCGT\\
$\Delta D_L/\hat{D}_L \simeq \hat{D}_L/(2.82\,\rm{Gpc})$ for LIGO + Virgo + AIGO + LCGT

\item {\it When isotropic emission is assumed, we find a large scatter
in distance measurement errors for all events, irrespective of network
and true distance.}  We find much less scatter when we assume a
beaming prior.  This is illustrated very clearly by the upper-right
panel of Fig.\ {\ref{fig:DeltaDLNSNS}}.  In that panel, we show the
scatter of distance measurement error versus true distance for the
LIGO, Virgo, AIGO detector network, comparing to the
Fisher-matrix-derived linear scaling trend found in DHHJ06.  For the
unbeamed case, our current results scatter around the linear trend;
for the beamed case, most events lie fairly close to the trend.  This
demonstrates starkly the failure of Fisher methods to estimate
distance accuracy, especially when we cannot set a beaming prior.

\item {\it Adding detectors to the network considerably increases the
number of detected binaries, but does not significantly improve the
accuracy with which those binaries are measured.}  The increase we see
in the number of detected binaries is particularly significant for
GW-SHB standard sirens.  For instance, an important application is
mapping out the posterior PDF for the Hubble constant, $H_0$.  As the
number of events increases, the resulting joint posterior PDF in $H_0$
will become increasingly well constrained.  Additional detectors also
increase the distance to which binaries can be detected.  This can be
seen in Fig.\ \ref{fig:DeltaDLNSNS}: for the LIGO and Virgo network,
our detected events extend to $\hat D_L \sim 600\,{\rm Mpc}$; the
larger networks all go somewhat beyond this.  Interestingly, networks
which include the AIGO detector seem to reach somewhat farther out.

\end{itemize}

It is perhaps disappointing that increasing the number of detectors
does not greatly improve measurement accuracy.  We believe this is due
to two effects.  First, a larger network tends to detect more weak
signals.  These additional binaries are poorly constrained. Second,
the principle limitation to distance measurement is the
$D_L$--$\cos\iota$ degeneracy.  A substantial improvement in distance
accuracy on individual events would require breaking this degeneracy.
We find that adding detectors does not do this, but the beaming prior
does.

\subsection{NS-BH binaries}

\begin{figure}
\centering 
\includegraphics[width=1.0\columnwidth]{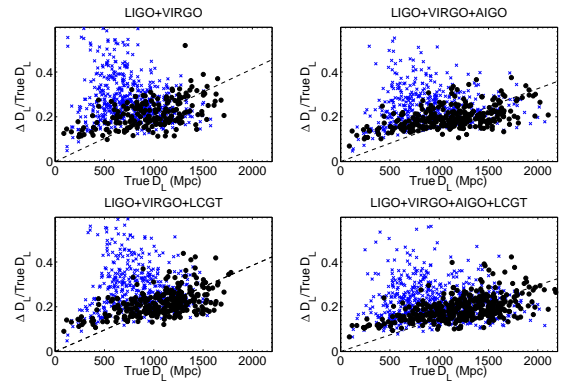}
\caption{Distance measurement errors versus true luminosity distance
for our sample of NS-BH binaries.  Colored crosses assume isotropic
emission; black points use our beaming prior.  The lower right-hand
panel shows the sample detected by our ``full'' network
(LIGO+Virgo+AIGO+LCGT).  Upper left is LIGO+Virgo; upper right is
LIGO+Virgo+AIGO; and lower left is LIGO+Virgo+LCGT. The dashed lines
show the linear best-fit to the beamed sample (see text for
expressions).}
\label{fig:DeltaDLNSBH}
\end{figure}

We now repeat the preceding analysis for six hundred detected NS-BH
binaries.  Figure~\ref{fig:DeltaDLNSBH} shows scatter plots of
measurement accuracies for unbeamed and beamed NS-BH binaries.  We
find similar trends to the NS-NS case:

\begin{itemize}

\item {\it The unbeamed and beamed samples separate into two distinct
distributions.}  Notice, however, that outliers exist in measurement
errors at high $D_L$ for several beamed events for all networks.  This
is not too surprising, given that we expect beamed sources at higher
luminosity distances and lower SNR.  Such events are more likely to
deviate from the linear relationship predicted by the Fisher matrix.

\item {\it We see substantial scatter in distance measurement,
particularly when isotropic emission is assumed}.  As with the NS-NS case, the
scatter is not as severe when we assume beaming, and in that case lies
fairly close to a linear trend, as would be predicted by a Fisher
matrix.  This trend is shallower in slope than for NS-NS binaries,
thanks to the larger mass of the system.

\item {\it We do not see substantial improvement in distance
measurement as we increase the detector network.}  As with NS-NS binaries,
adding detectors increases the range of the network; AIGO appears
to particularly add events at large $\hat D_L$ (for both the isotropic
and beamed samples).  However, adding detectors does not break the
fundamental $D_L$--$\cos\iota$ degeneracy, and doesn't improve errors.  From our
full posterior PDFs, we find absolute errors of $0.1$--$0.3$ in $\cos\iota$,
which is very similar to the NS-NS case.

\item {\it Beamed subsample scalings.} The linear scalings
for our beamed subsample are:\\
$\Delta D_L/\hat{D}_L \simeq \hat{D}_L/(4.83\,\rm{Gpc})$ for LIGO + Virgo\\
$\Delta D_L/\hat{D}_L \simeq \hat{D}_L/(6.14\,\rm{Gpc})$ for LIGO + Virgo + AIGO\\
$\Delta D_L/\hat{D}_L \simeq \hat{D}_L/(5.20\,\rm{Gpc})$ for LIGO + Virgo + LCGT\\
$\Delta D_L/\hat{D}_L \simeq \hat{D}_L/(6.76\,\rm{Gpc})$ for LIGO + Virgo + AIGO + LCGT\\

\end{itemize}

\section{Summary discussion}
\label{sec:summary}

In this analysis we have studied how well GWs can be used to measure
luminosity distance, under the assumption that binary inspiral is
associated with (at least some) short-hard gamma ray bursts.  We examine two
plausible compact binary SHB progenitors, and a variety of plausible
detector networks.  We emphasize that {\it we assume sky position is
known}.  We build on the previous study of DHHJ06, which used the
so-called Gaussian approximation of the posterior PDF.  This
approximation works well for large SNR, but the limits of its validity are
poorly understood.  In particular, since the SNR of events measured by ground-based
detectors is likely to be of order 10, the Gaussian limit may be inapplicable. 
We examine the posterior PDF for the parameters of observed
events using Markov-Chain Monte-Carlo techniques, which do not rely on
this approximation.  We also introduce a well-defined noise-averaged
posterior PDF that does not depend solely on a particular noise
instance. Such a quantity is useful to predict how well a detector
should be able to measure the properties of a source.

We find that the Gaussian approximation substantially underestimates
distance measurement errors.  We also find that the main limitation
for individual standard siren measurements is the strong degeneracy
between distance to the binary and the binary's inclination to the line of
sight; similar discussion of this issue is given in a recent analysis
by \cite{ajithbose09}.  Adding detectors to a network only slightly
improves distance measurement for a given single event.  When we
assume that the SHB is isotropic (so that we cannot infer anything
about the source's inclination from the burst), we find that Fisher
matrix estimates of distance errors are very inaccurate.  Our
distributions show large scatter about the Fisher-based predictions.

The situation improves dramatically if we assume that SHBs are
collimated, thereby giving us a prior on the orientation of the
progenitor binary.  By assuming that SHBs are preferentially emitted
into an opening angle of roughly $25^\circ$, we find that the
distance--inclination correlation is substantially broken.  The Fisher
matrix estimates are then much more reasonable, giving a good sense of
the trend with which distances are determined (albeit with a moderate
scatter about that trend).  This illustrates the importance of
incorporating prior knowledge, at least for individual measurements.

Our distance measurement results are summarized by
Fig.\ {\ref{fig:DeltaDLNSNS}} (for NS-NS SHB progenitors) and
Fig.\ {\ref{fig:DeltaDLNSBH}} (for NS-BH).  Assuming isotropy, we find
the distance to NS-NS binaries is measured with a fractional error of
roughly $20$--$60$\%, with most events in our distribution clustered
near $20$--$30$\%.  Beaming improves this by roughly a factor of two,
and eliminates much of the high error tail from our sample.  NS-BH
events are measured somewhat more accurately: the distribution of
fractional distance errors runs from roughly $15$--$50$\%, with most
events clustered near $15$--$25$\%.  Beaming again gives roughly a
factor of two improvement, elimating most of the high error tail.

It is worth emphasizing that these results describe the outcome of
{\it individual siren measurements}.  When these measurements are used
as cosmological probes, we will be interested in constructing the
joint distribution, following observation of $N$ GW-SHB events.
Indeed, preliminary studies show that our ability to constrain $H_0$
improves dramatically as the number of measured binaries is
increased.  In our most pessimistic scenario (the SHB is assumed to be
a NS-NS binary, with no prior on inclination, and measured by the
baseline LIGO-Virgo network), we find that $H_0$ can be measured with
$\sim 13\%$ fractional error with $N = 4$, improving to $\sim 5\%$ for
$N = 15$.  This is because multiple measurements allow us to sample
the inclination distribution, and thus average out the bias
introduced by the tendency to overestimate distance for edge-on binaries,
and underestimate it for face-on binaries.  Details of this analysis
will be presented in a followup paper.

Increasing the number of measured events will thus be crucial for
making cosmologically interesting measurements.  To this end, it is
important to note that increasing the number of detectors in our
network enables a considerable increase in the number of detected
binaries.  This is due to increases in both the sky coverage and in the
total detection volume.  Going from a network which includes all four
detectors (LIGO, Virgo, AIGO, and LCGT) to our baseline network of
just LIGO and Virgo entails a $\sim$~50\% reduction in the number of
detected binaries. Eliminating just one of the proposed detectors
(AIGO or LCGT) leaves us with $\sim$~75\% of the original detected
sample.

Aside from exploring the cosmological consequences, several other
issues merit careful future analysis.  One general result is
the importance that priors have on the posterior PDF.  We plan to
examine this in some detail, identifying the parameters which particularly
influence the final result, and which uncertainties can be ascribed to
an inability to set relevant priors.  Another issue is the importance of
systematic errors in these models.  We have used the second-post-Newtonian
description of a binary's GWs in this analysis, and have ignored all
but the leading quadrupole harmonic of the waves (the ``restricted''
post-Newtonian waveform).  Our suspicion is that a more complete
post-Newtonian description of the phase would have little impact on
our results, since such effects won't impact the $D_L$--$\cos\iota$
degeneracy.  In principle, including additional (non-quadrupole)
harmonics could have an impact, since these other
harmonics encode different information about the inclination angle
$\iota$.  In practice, we expect that they won't have much effect on
GW-SHB measurements, since these harmonics are measured with very low
SNR (the next strongest harmonic is roughly a factor of 10 smaller in
amplitude than the quadrupole).

As discussed previously, we confine our analysis to the inspiral.
Inspiral waves are terminated at the innermost stable circular orbit
frequency, $f_{\rm ISCO}=(6^{3/2} \pi M_z)$.  For NS-NS binaries,
$f_{\rm ISCO} \simeq 1600\,{\rm Hz}$.  At this frequency, detectors
have fairly poor sensitivity, so we are confident that terminating the
waves has little impact on our NS-NS results.  However, for our
assumed NS-BH binaries, $f_{\rm ISCO} \simeq 400\,{\rm Hz}$.
Detectors have good sensitivity in this band, so it may be quite
important to improve our model for the waves' termination in this
case.

Perhaps the most important follow-up would be to include the impact of
spin.  Although the impact of neutron star spin is likely to be small,
it may not be negligible; and, for NS-BH systems, the impact of the
black hole's spin is likely to be significant.  Spin induces
precession which makes the orbit's orientation, $\bf{\hat L}$,
dynamical.  That makes the observed inclination dynamical, which can
break the $D_L$--$\cos\iota$ degeneracy.  In other words, with spin
precession the source's orbital dynamics may break this degeneracy.
Van der Sluys et al.\ (2008) have already shown that spin precession
physics can improve the ability of ground-based detectors to determine
a source's position on the sky.  We are confident that a similar
analysis which assumes known sky position will find that measurements
of source distance and inclination can likewise be improved.

\acknowledgments

It is a pleasure to acknowledge useful discussions with K.\ G.\ Arun,
Yoicho Aso, Duncan Brown, Curt Cutler, Jean-Michel D{\'e}sert,
Alexander Dietz, L.\ Samuel Finn, Derek Fox, \'Eanna Flanagan, Zhiqi
Huang, Ryan Lang, Antony Lewis, Ilya Mandel, Nergis Mavalvala,
Szabolcs M\'arka, Phil Marshall, Cole Miller, Peng Oh, Ed Porter,
Alexander Shirokov, David Shoemaker, and Pascal Vaudrevange.  We are
grateful to Neil Cornish in particular for early guidance on the
development of our MCMC code, to Michele Vallisneri for careful
reading of the manuscript, and to Phil Marshall for his detailed
comments on the ensemble averaged likelihood function. We also are
grateful for the hospitality of the Kavli Institute for Theoretical
Physics at UC Santa Barbara, and to the Aspen Center for Physics,
where portions of the work described here were formulated.
Computations were performed using the Sunnyvale computing cluster at
the Canadian Institute for Theoretical Astrophysics, which is funded
by the Canadian Foundation for Innovation.  SAH is supported by NSF
Grant PHY-0449884, and the MIT Class of 1956 Career Development Fund.
He gratefully acknowledges the support of the Adam J.\ Burgasser Chair
in Astrophysics.

\bibliography{sirens}

\end{document}